\documentclass[apj]{emulateapj}
\usepackage[active]{srcltx}
\usepackage{color}
\usepackage{lscape}

\shorttitle{Probing the GC-LMXB connection in NGC 1399}
\shortauthors{Paolillo et al.}

\begin{document}

\title{Probing the GC-LMXB connection in NGC 1399: a wide-field study with HST and \textit{Chandra}\altaffilmark{*}.}

\author{Maurizio Paolillo\altaffilmark{1,2},Thomas H. Puzia\altaffilmark{3,9}, Paul Goudfrooij\altaffilmark{4}, Stephen E. Zepf\altaffilmark{5},Thomas J. Maccarone\altaffilmark{6}, Arunav Kundu\altaffilmark{7}, Giuseppina Fabbiano\altaffilmark{8}, Lorella Angelini\altaffilmark{10}
}

\altaffiltext{1}{Dept.of Physical Sciences, University of Napoli Federico II,via Cinthia 9, 80126, Italy}
\email{paolillo@na.infn.it}
\altaffiltext{2}{INFN - Napoli Unit, Dept. of Physical Sciences, via Cinthia 9, 80126, Napoli, Italy}
\altaffiltext{3}{Herzberg Institute of Astrophysics, 5071 West Saanich Road, Victoria, BC V9E 2E7, Canada}
\altaffiltext{4}{Space Telescope Science Institute, 3700 San Martin Drive, Baltimore, MD 21218, USA}
\altaffiltext{5}{Dept.of Physics and Astronomy, Michigan State University, East Lansing, MI 48824, USA}
\altaffiltext{6}{School of Physics and Astronomy, University of Southampton, Southampton SO17 1BJ, United Kingdom}
\altaffiltext{7}{Eureka Scientific, 2452 Delmer St., Oakland, CA 94602, USA}
\altaffiltext{8}{Harvard-Smithsonian Center for Astrophysics, 60 Garden St., Cambridge, MA 02138, USA}
\altaffiltext{9}{Dept.of Astronomy and Astrophysics, Pontificia Universidad Cat\'olica de Chile, 7820436 Macul, Santiago, Chile}
\altaffiltext{10}{Laboratory for High Energy Astrophysics, NASA Goddard Space Flight Center, Code 660, Greenbelt, MD 20771, USA}

\altaffiltext{*}{Based on observations with the NASA/ESA {\it Hubble
    Space Telescope}, obtained at the Space Telescope Science
  Institute, which is operated by the Association of Universities for
  Research in Astronomy, Inc., under NASA contract NAS5-26555}

\begin{abstract}
We present a wide field study of the Globular
Clusters/Low Mass X-ray Binary (LMXB) connection in the giant elliptical NGC1399.
The large FOV of the ACS/WFC, combined with the HST and \textit{Chandra} high resolution, 
allow us to constrain the LMXB formation scenarios in elliptical galaxies. 
We confirm that NGC1399 has the highest LMXB fraction in GCs of all
nearby elliptical galaxies studied so far, even though the exact value depends on galactocentric distance due to the interplay of a differential GC vs galaxy light distribution and the GC color dependence. In fact LMXBs are preferentially hosted by bright, red GCs out to $>5\, R_{\rm eff}$ of the galaxy light.
The finding that GC hosting LMXBs follow the radial distribution of their parent GC population, 
argues against the hypothesis that the external dynamical influence of the galaxy affects LMXB formation in GCs.
On the other hand field LMXBs closely match the host galaxy light, thus indicating that they are originally formed
in situ and not inside GCs. We measure GC structural parameters, finding that the LMXB formation likelihood is influenced independently by mass, metallicity and GCs structural parameters. In particular the GC central density plays a major role in predicting which GC host accreting binaries.
Finally our analysis shows that LMXBs in GCs are marginally brighter than those in the field, and in particular the only color-confirmed GC with $L_X>10^{39}$ erg s$^{-1}$ shows no variability, which may indicate
a superposition of multiple LMXBs in these systems. 
\end{abstract}


\section{Introduction}

A significant contribution from accreting binary stars to the total
X-ray emission of early-type galaxies has been predicted for a long
time, using the X-ray/optical luminosity ratio and spectral energy
distribution \citep[see][and references therein]{Fabbiano89} as primary indicators, long
before the majority of X-ray sources could be resolved individually. 
With the launch of {\it Chandra}, with its
sub-arcsecond spatial resolution, tens to hundreds of low-mass X-ray
binaries (LMXB) were discovered in nearby ellipticals, a large number of which
are residing in Globular Clusters (GC), with a complex dependence
on the properties of the host galaxy and of the GC population.

Several studies have shown that while on average a few
percent ($\sim\!5\%$) of GCs host LMXBs, the fraction of LMXBs residing
in GCs varies from $10\!-\!20\%$ in late-type galaxies and
reaches up to $\sim\!70\%$ in cD galaxies, depending on the
morphological type of the galaxy and on the GC specific frequency
\citep[see review in][]{Fabbiano06}. It was also observed that LMXBs reside
preferentially in bright GCs \citep{Angelini01, Kundu02, Sarazin03,
Kim06, Kundu07, Sivakoff07}, as expected if dynamical interactions
favor binary formation in dense environments \citep{Clark75, white02, Pooley03,
Verbunt05}. More puzzling is the dependence of the probability of
finding LMXBs on the GC color. Recent studies indicate that red (old, metal-rich) GCs
are $\sim\!3$ times more likely to host LMXBs than blue
(young, metal-poor) ones \citep{Angelini01, Kundu02, Sarazin03,
Jordan04, Kim06, Kundu07, Sivakoff07}.

The spatial distribution of LMXBs is also debated: while some studies
find that the spatial distribution of GC-LMXBs is more extended than
field-LMXBs \citep{Kim06,Kundu07}, others do not observe such a difference
\citep[e.g][]{Humph08}. The issue is further complicated
by the fact that for a proper comparison with the distribution of host
GCs, the samples must be split according to GC colors \citep[see][ and 
references therein]{Fabbiano06}.

Constraining these observables is crucial for discriminating among LMXB
formation models. For instance, irradiation-induced winds \citep{MKZ04},
magnetic breaking \citep{Ivanova06} or IMF variations
(\citealp{Grindlay87}, \citealp[also see][]{Jordan04}) can explain the
LMXB formation likelihood as a function of the host GC color in terms of
a metallicity effect, while other dynamical models \citep[e.g][]{Kim06,
Jordan07a, Sivakoff07} suggest that this color dependence may reflect
the higher LMXB formation efficiency in more centrally-concentrated red
GCs.

Two main observational problems affect our current ability to understand
the importance of external dynamical factors governing the LMXB
formation in GCs. First, most studies of the LMXB/GC connection have
been restricted to the central regions of nearby ellipticals, due to the
limited field of view (FOV) surveyed by space observatories, i.e.~HST
and {\it Chandra}, whose high spatial resolution is required to minimize
the positional uncertainties and reduce the background contamination.
This has prevented detailed studies of how the distance from the galaxy
center and orbital motions affect the LMXB formation efficiency in GCs.
Furthermore, since the radial distributions of red and blue GC are known
to be different, with the former being more centrally-concentrated than
the latter, a restricted FOV introduces systematic sample selection
biases. The few wide-field studies that have tried to address this issue from the ground (see $\S$ \ref{discussion}), did not yield conclusive results due to the large background contamination.

Second, until a few years ago little was known about GC sizes outside
the Local Group \citep[e.g.][]{kundu98, kundu99, puzia99, puzia00}, due
to angular resolution and FOV limits of earlier generations of HST
instruments. The HST/ACS camera with its high spatial resolution and
efficiency has more recently allowed us to resolve GC sizes in many
nearby massive ellipticals down to a few pc \cite[e.g.][]{Jordan05,
Jordan07a, Jordan07b, Sivakoff07}. Again, these studies are mainly
limited to the central regions of the galaxy and to the brightest GCs.

In this context we initiated a project to perform a wide-field, high
spatial-resolution study of GCs and LMXBs in one of the closest giant
ellipticals, NGC~1399, with a very rich GC system. Located at about 20 Mpc
distance \citep[$D=20.13\pm0.4$ Mpc, see][]{dunn06}, this galaxy is near enough to resolve GC sizes
with ACS, while distant enough to sample efficiently the GC distribution
out to large galactocentric radii. Furthermore this object is believed
to have one of the highest fractions of LMXBs residing in GCs
\citep{Angelini01,Kim06}, thus providing a large sample of field- and
GC-LMXBs.

In a parallel article to this one \cite[][ in prep. - hereafter P11]{PaperI} we
present the HST/ACS data and structural parameter analysis. Here we focus 
on the GC/LMXB connection and discusses its dependence on galactocentric distance and
GC properties.

\section{Observations}
\subsection{The HST/ACS data}
\label{opt_data_sect}
A detailed description of the HST data and source catalogs are given in
P11. Here we briefly summarize the properties of the optical dataset
for the sake of completeness.
\begin{figure}[t]
\includegraphics[width=8.5cm]{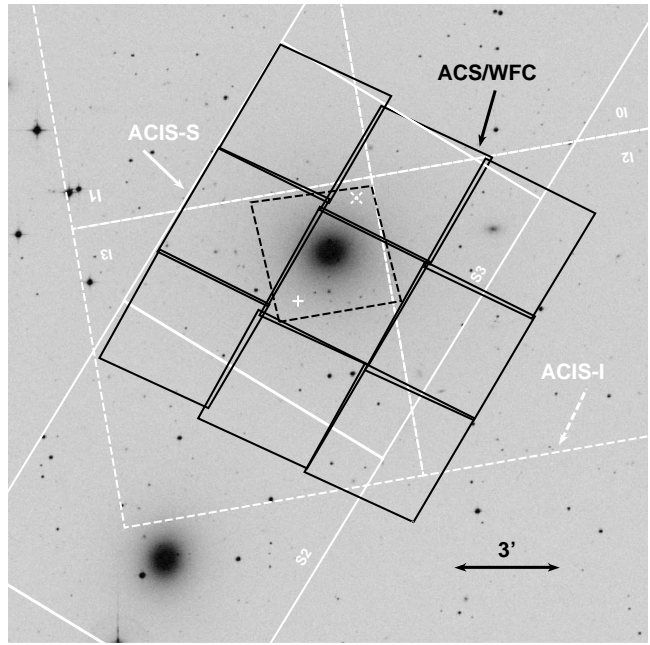}
\caption{Illustration of the 3x3 mosaic of our ACS observations (black
solid lines) overplotted on a DSS-2 image of the NGC 1399 region. Also
shown (black dashed polygon) is the FOV of the archival ACS observations
covering the galaxy center. The white solid (dashed) lines outline the
FOV of the {\it Chandra} ACIS-S (ACIS-I) chips with the aimpoint marked
by the '+' ('X') symbol.}
\label{fov}
\end{figure}
The optical data were taken with the {\it Advanced Camera for Surveys}
\citep[ACS,][]{ford03} onboard the {\it Hubble Space Telescope}
(GO-10129), in the F606W filter with a total integration time of 2108
seconds per pointing. The observations were arranged in a $3\times3$ ACS
mosaic as illustrated in Figure~\ref{fov}. The individual observations
were dithered to allow sub-pixel sampling of the ACS PSF and combined
into a single image using the MultiDrizzle routine 
\citep{koekemoer02}. The final scale\footnote{At NGC~1399 distance 1 pix
= 2.93 pc; 1\arcsec\ = 97.7 pc.} of the images is 0.03\arcsec/pix and
provides a super-Nyquist sampling of the stellar point spread function
with a full width at half maximum (FWHM) of $\sim\!0.09\arcsec$.

\begin{deluxetable}{lcc}[t!]
\centering
\tabletypesize{\scriptsize}
\tablecaption{Photometric selection criteria for GC candidates\label{col_sel}}
\tablehead{\colhead{} & \colhead{blue GCs} & \colhead{red GCs}}
\startdata
Ground-based & $T1<23$& $T1<23$\\
data & $1.0\leq C\!-\!T1<1.65$ & $1.65\leq C\!-\!T1<2.2$ \\\\
HST data & $z<22.5$ & $z<22.5$ \\
 & $1.3\leq g\!-\!z<1.9$ & $1.9\leq g\!-\!z<2.5$ \\
\enddata
\end{deluxetable}

To maximize the overlap with other observations (e.g. X-ray imaging and
ground-based spectroscopy) the entire mosaic was centered on the
coordinates: RA (J2000)~$\!=\!03^{\rm h} 38^{\rm m} 28.62^{\rm s}$ and
Dec (J2000)~$\!=\!-35^{\rm o} 28\arcmin\ 18.9\arcsec$. The field of view
of the ACS mosaic covers $\sim\!100$ square arcminutes and extends out
to a maximum projected galactocentric distance of $\sim\!50$ kpc with
respect to NGC~1399, i.e.~$\sim\!5.2$ effective radii of the diffuse
galaxy light \citep{RC3} and $\sim\!4.9$ core radii of the globular
cluster system density profile \citep{Schuberth10}.

Source catalogs were generated with SExtractor, using the appropriate
weight maps produced by the Multidrizzle procedure, requiring a minimum
area of 20 pixels and total S/N$>7$. The catalog astrometric solution
was registered using the USNO-B1
catalog\footnote{http://tdc-www.harvard.edu/software/catalogs/ub1.html}
as a reference frame. Bright, unsaturated stars were identified in all
ACS frames and matched with USNO-B1 sources obtaining a final accuracy
of 0.2\arcsec\ r.m.s.

Brightness estimates $m_V(F606W)$ in the STMAG photometric system 
for all detected sources were derived from isophotal magnitudes measured by
SExtractor. While this approach is not optimal for resolved GCs, our
accuracy is appropriate for the present work. Refined photometry for all
GC candidates is computed in P11 in the VEGAMAG system, and 
compares well with our measurement, yielding an average conversion 
factor of $m_{\rm STMAG}-m_{\rm VEGA}=0.16$ mag at $m_V=23.5$ 
mag with a scatter of 0.04 mag.

\begin{figure}[t]
\includegraphics[width=8.5cm]{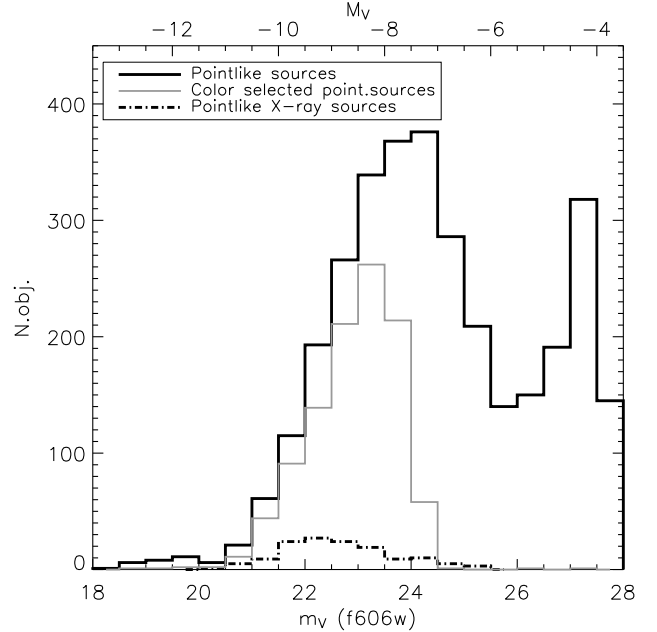}
\caption{Luminosity distributions of optical and X-ray 
point-like sources (e.g. with stellarity index $<0.9$) within the HST FOV. Also shown is the color-selected (Table \ref{col_sel}) subsample of pointlike sources.}
\label{maghisto}
\end{figure}

Since no complete color catalog was available for the whole field, GC
candidates were selected based on magnitude and morphological
classification, choosing sources with SExtractor stellarity index $\geq 0.9$
and magnitude $m_V\!<\!26$ mag in order to exclude extended sources and
compact background galaxies. The magnitude distribution of all point-like
sources in our fields is shown in Figure~\ref{maghisto}. The distribution
closely follows the GC luminosity function down to $m_V\!\lesssim\! 26$
mag; at fainter magnitudes background unresolved sources dominate the
number counts.

To include optical color information in our analysis we use the
\citet{Bassino06} $C-T1$ ground-based GC catalog\footnote{The original 
filters used by \citet{Bassino06} were Washington $C$ and Harris $R$. 
The standard stars were instead taken in Washington $T1$ and colors $C-T1$. 
Since $R$ and $T1$ are almost identical (difference 0.02 mag) we use $T1$ 
throughout the paper.},
which contains data for
$\sim\!50\%$ of our GC candidates within the HST FOV. 
Since the ground-based catalog is incomplete within
40\arcsec\ from the NGC~1399 center due to galaxy light contamination, we
included in our analysis the HST/ACS $g-z$ color catalog from \cite{Kundu05}\footnote{In this catalog a uniform aperture correction was used for all sources.}
which provides colors for $\sim 90\%$ of GC candidates in the central region of the galaxy (see Figure \ref{fov}). GCs were divided into blue and red populations as described in Table~\ref{col_sel}; note that the magnitude limit is chosen to ensure an approximately uniform completeness across the whole color and galactocentric distance range (Figure \ref{col_mag}, also see \citealt{Bassino06}).

In order to confirm the reliability of our GC selection method, based on single-band F606W photometry,
and compare it with the color-selection usually adopted in the literature,
we measure the fraction of GC candidates within the subset of sources with color information.
Assuming that bona-fide GCs are represented by sources within the color ranges presented in Table \ref{col_sel},
we derive two different estimates: i) for the central region covered by the more accurate $g$ and $z$ HST photometry and,
ii) for the entire field covered by the ground based $C$ and $T1$ data.
Within the central region $80\%$ of our GC candidates (within $m_V<26$ by definition) are consistent with the $1.3\leq g-z<2.5$ color cut; restricting the analisis to the bright subsample with $z<22.5$, used in the following sections to study the red and blue sub-populations, this number increases to $92\%$. Using the $C-T1$ photometry instead, which extends over the whole HST mosaic we find that $82\%$ of the GC candidates are consistent with the $1.0\leq C\!-\!T1<2.2$ color and $T1<23$ magnitude cuts.\footnote{The completeness of our GC candidate sample \textit{with respect to the entire GC population} will be obviously lower. For instance, assuming that our GC candidates follow a lognormal distribution as suggested by Figure \ref{maghisto}, we calculate that our $m_V<26$ cut removes about 5\% of the entire GC population, thus resulting in a 76\% competeness level. The estimates based on $C-T1$ selection however, must be regarded as a lower limit, since using our F606W single-band HST data we find that $\sim 10\%$ of the sources which have $C-T1$ color consistent with GCs, are resolved as extended background galaxies.}
On the other hand, $\sim 4\%$ and $\sim 9\%$ of the GC candidates have respectively $g-z$ and $C-T1$ colors outside the allowed range as given in Table \ref{col_sel}.
We point out that using our stellarity selection criteria, we are able to effectively remove background galaxies since the fraction of such contaminants, which is expected to increase at large radii, varies by only by a few percent (from $7\%$ to $\sim 10\%$) across our entire FOV.
Also note that in the following Sections we treat the $C-T1$ and $g-z$ subsamples separately when the different completeness levels may affect our conclusions.

\begin{figure}[t]
\includegraphics[width=8.5cm]{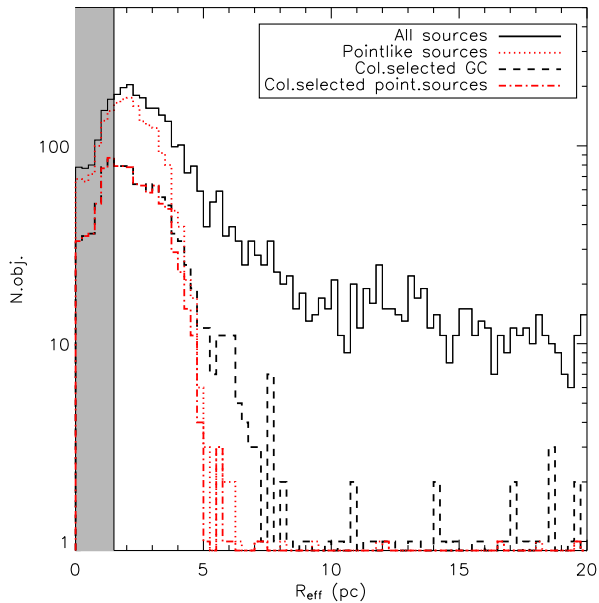}
\caption{Half-light radius distribution for the entire ACS optical catalog (solid line), compared to our GC candidates (dotted line). Restricting the sample to color-confirmed GCs (dashed and dot-dashed lines) shows that our selection criteria misses only very extended GCs with $R_{\rm eff}>5$ pc. The shaded region highlights the region where our size measurement are poorly constrained (see $\S$ \ref{struct_parms}).}
\label{size_distro}
\end{figure}

To test whether very extended GCs are misclassified by our selection criteria,  we
estimated the completeness of our bona-fide GC sample as a function of GC size for
the subset of optical sources with measured structural parameters (see $\S$ \ref{struct_parms});
in Figure \ref{size_distro} we show the effective radius distribution of the GC candidates samples,
finding that our completeness drops below $50\%$ only for $R_{\rm eff}>5$ pc, with respect to color selected GCs. 
We also verified that relaxing the Stellarity index criterium does not increase the completeness for large ($R_{\rm eff}>5$ pc) GCs since these are fully resolved on our HST images, while increasing significantly the contamination level. 

\subsection{{\it Chandra} X-ray data}
\label{Chandra_data}
The X-ray data were retrieved from the {\it Chandra} public
archive\footnote{http://cxc.harvard.edu}. We selected observations \#319
and \#1472, i.e. two imaging datasets with the long exposure times where
NGC~1399 lies close to the ACIS aimpoint, for a total exposure time of
$\sim 100$ ks\footnote{The additional observations available in the archive 
would not significantly increase the S/N ratio, while complicating the data
analysis process and increasing the impact of PSF variation systematics.}.

\begin{figure}[t!]
\includegraphics[width=8.5cm]{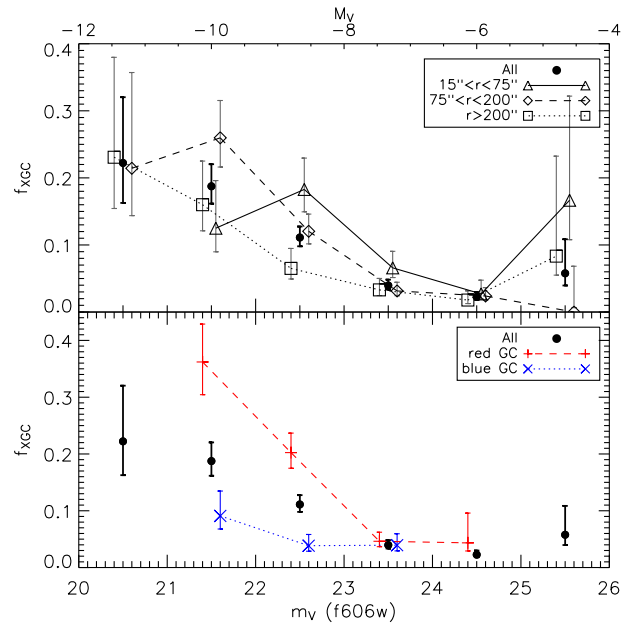}
\caption{Fraction of GCs hosting LMXBs as a function of luminosity. The two panels show parametrizations of those fractions by galactocentric distance ({\it upper panel}) or GC color ({\it lower panel}). The top abscissa gives the absolute magnitudes for the assumed Fornax distance of 20.13 Mpc \citep{dunn06}. The points are slightly shifted for visualization purposes.}
\label{LMXB_frac}
\end{figure}

The X-ray data were reduced with the CIAO software, extracting 
standard-grade events after applying bad pixel mask and afterglow corrections.
The final exposure times, after removing high background periods, are
shown in Table \ref{table_obs}. To maximize the astrometric accuracy of
the observations particular care was taken to correct for known
offsets\footnote{http://cxc.harvard.edu/cal/ASPECT/celmon/} and to
reproject the aspect solution and the event files of the individual
observations using the NGC~1399 centroid as a reference point. In both
cases the total offset was $<\!1\arcsec$. To minimize the uncertainties
due to completeness variations over the FOV, we limited our analysis to
the region in common to ACIS-I, ACIS-S and HST/ACS (see Figure
\ref{fov}). The subsequent analysis is thus limited to this overlap
region.

\begin{figure*}[t!]
\centering{\includegraphics[width=15cm]{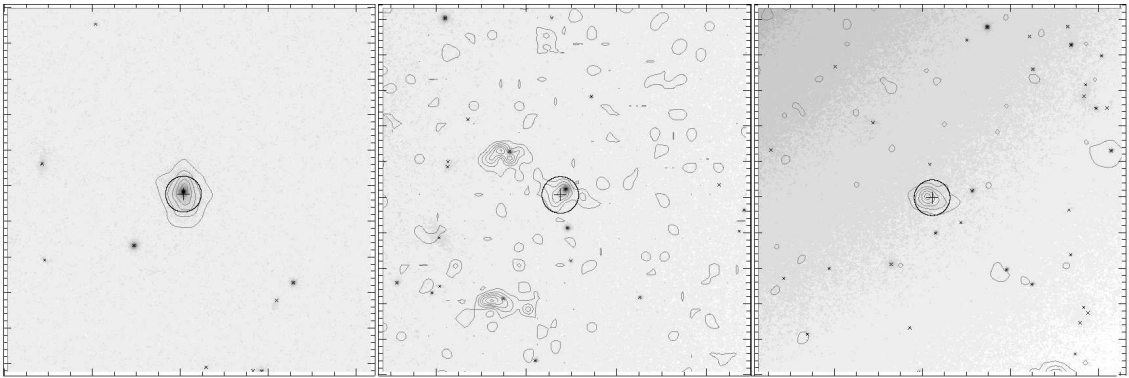}}
\caption{X-ray contours overlayed on ACS images: a very precise match 
(left), matched within 1\arcsec\ (center) and X-ray source with no optical 
counterpart (right). The cross marks the X-ray centroid while the circle is 
our matching radius.}
\label{x-opt_match}
\end{figure*}

\begin{deluxetable*}{cccccc}[ht!]
\tabletypesize{\scriptsize}
\tablecaption{Journal of {\it Chandra} X-ray Observations\label{table_obs}}
\tablehead{\colhead{Obs.} & \colhead{Detector} & \colhead{Date} & \colhead{RA(J2000)} & \colhead{DEC(J2000)} & 
\colhead{$T_{\rm exp}$}
}
\startdata
\#319 & ACIS-S & 2000-01-18 & 03h38m29.4s & $-35^o$27\arcmin 00.4\arcsec & 56 ks\\
\#1472 & ACIS-I & 2003-05-26 & 03h38m25.6s & $-35^o$25\arcmin 42.6\arcsec & 45 ks\\
\enddata
\end{deluxetable*}

The {\sc wavdetect} algorithm \citep{Freeman02} was used to obtain a
preliminary source catalog for each dataset, using detection pixel
scales of 1,2,4,8,16 and a significance threshold of $10^{-6}$. The
registered observations were then aligned using the {\sc wavedetect}
catalogs with a residual positional uncertainty of 0.3\arcsec\ r.m.s.
and merged. Exposure maps were generated for both the individual
observations and the merged dataset, and a merged source catalog was
generated with {\sc wavdetect}. In generating the X-ray catalogs the
detection algorithm was run on both the whole $0.3\!-\!8$ keV energy
band and on the $0.3\!-\!1$, $1\!-\!2$, $2\!-\!8$ narrow bands, finding
$\sim\!12$ additional sources detected only in one of the narrow bands
out of a total of 230 X-ray sources. For comparison with the literature,
we note that 38 sources were not detected in the \#319 dataset
\cite[cf.][]{Angelini01} and half of these were only detected in the
merged dataset.

We used the ACIS Extract software\footnote{The {\em ACIS Extract} software
package and User's Guide documents are available for download at
\url{http://www.astro.psu.edu/xray/acis/acis\_analysis.html}.}
\cite[AE,][]{Broos10} to account for the variable PSF in the two X-ray
datasets, as well as to improve the positional accuracy. AE uses library
templates of the ACIS PSF to model the actual observed PSF for each
observation, as well as for the composite one, allowing to derive source
positions, properties and detection likelihood. We feed AE with the final
source list derived from combining both individual and merged {\sc
wavdetect} catalogs. Each source was then inspected individually to check
the position accuracy and to remove spurious objects due to poor data
quality (very faint sources, high X-ray background in the field center,
overlapping sources etc.), resulting in the removal of 3 objects, in
agreement with the contamination expected based on the {\sc wavdetect}
significance threshold.

The AE software provides three different position estimates: input
catalog, data centroid and correlation peak; our tests (as well as the AE
manual) suggest that the `data centroid' is the best estimate of the
source position. Furthermore, we found that it is quite consistent with
the {\sc wavdetect} positions within $\sim\!0.1\arcsec$ except for the
faintest objects and those farthest from the aimpoint where the difference
can reach 0.5\arcsec.~For three sources we decided to adopt the
correlation peak which seemed a more reliable estimate after visual
inspection; in any case, the difference between these estimates was always
$<\!1\arcsec$.
Finally the source catalogs were registered to the USNO-B1 reference
frame using bright X-ray sources with optical counterparts (mostly GCs).
The final accuracy of the X-ray catalog is 0.33\arcsec\ with a maximum
systematic offset of 0.6\arcsec. 

The properties of the 230 X-ray sources in the composite {\it Chandra}/HST FOV are summarized in Table \ref{tbl-1}. 

\section{Analysis}

\subsection{Matching Optical and X-ray Data}
\label{matching}

\begin{figure*}[ht!]
\centering
\includegraphics[width=8.5cm]{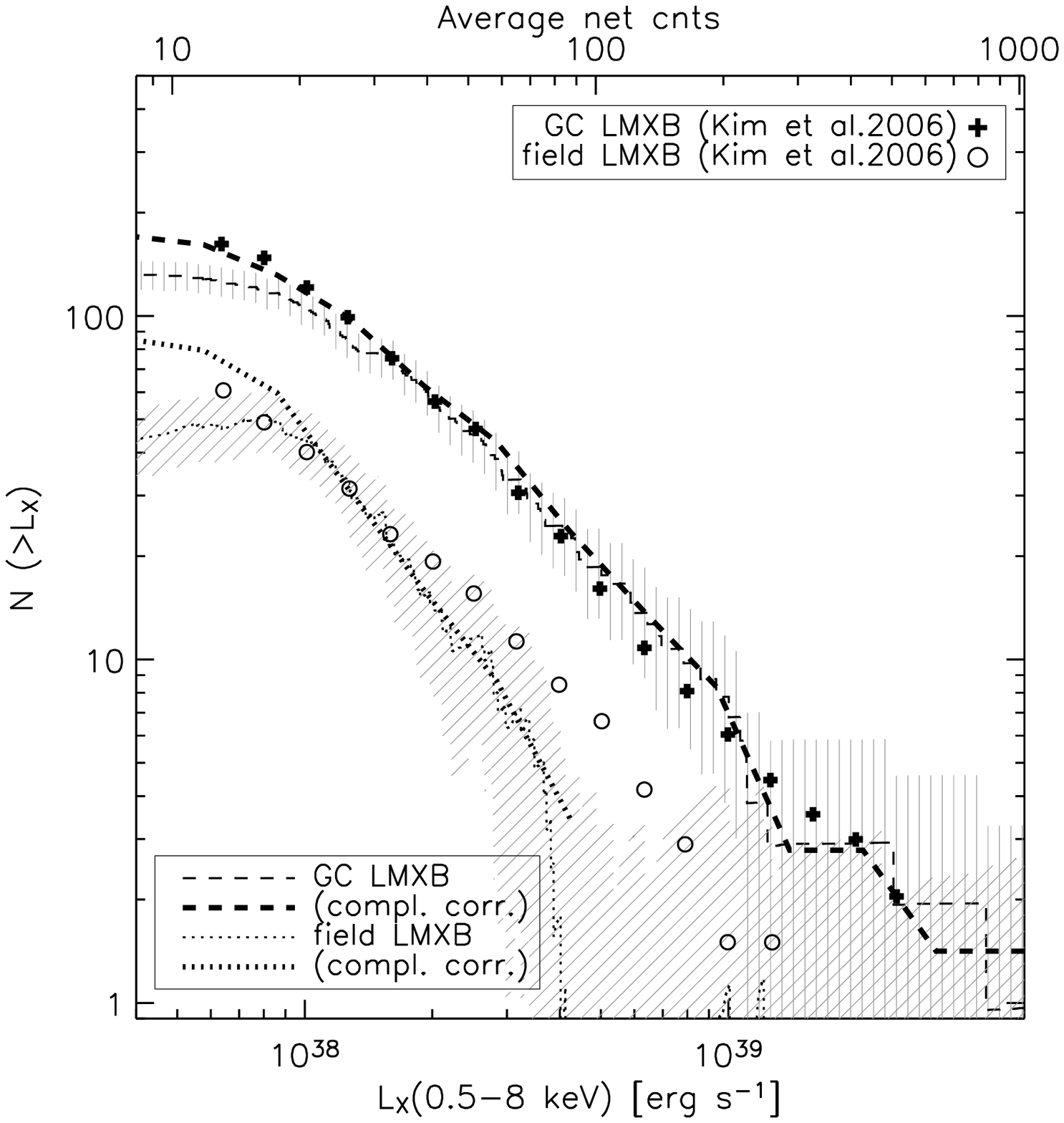}
\includegraphics[width=8.5cm]{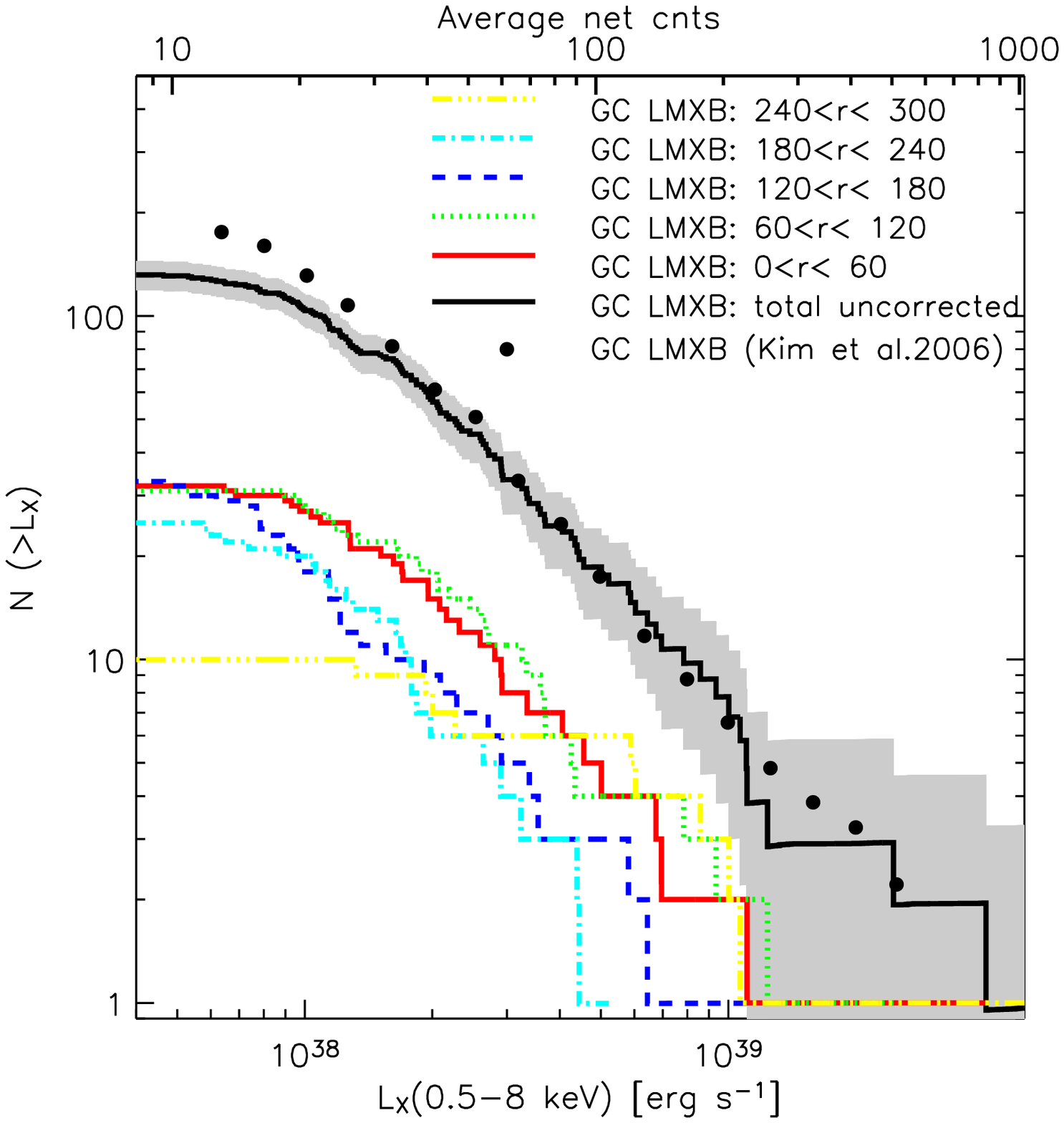}
\caption{{\it Left panel:} Cumulative X-ray Luminosity Function of GC and
field LMXB. We observe a good agreement with \citet[ arbitrarily
re-normalized]{Kim06} for GC-LMXB, while our LF for field LMXB is marginally steeper (see discussion in text). The shaded region represent $1\sigma$ errors. {\it Right
panel:} X-ray luminosity function in different radial bins.}
\label{LF}
\end{figure*}

To match the optical and X-ray sources we used the final catalogs
registered on the USNO-B1 reference frame described in previous sections.
Since the accuracy of the optical catalogs is $\sigma_{pos}^{opt}\simeq
0.2\arcsec$ while the X-ray one has $\sigma_{pos}^{X}\simeq 0.33\arcsec$,
we adopted a conservative $2.5\sigma$ matching radius of 1\arcsec.
Furthermore four sources, lying within the central 15\arcsec\ of the galaxy, were
excluded from the analysis because of the strong galaxy light
contribution, both in optical and X-rays. Our matching algorithm thus
yields 164 X-ray sources with optical counterparts, out of which 136 are
matched with GC candidates. Note that 75\% of these objects are matched
within 0.5\arcsec.
When multiple optical counterparts (usually 2) were present, i.e. for $\sim 14\%$ of the sample, we choose the closest one as the most likely match; however we verified that our conclusions do not change if we exclude such sources from our sample.
While the matching accuracy is expected to depend on the galactocentric distance, mainly due to the larger {\it Chandra} PSF toward the outskirts of the ACIS FOV, we verified that even at large radii ($>2$ arcmin) doubling the matching radius results in a mild ($\sim 10\%$) gain in the number of matched sources, while increasing by $30\%$ the contamination level (see below). Our conservative choice thus minimizes the contamination at the cost of losing some of the fainter X-ray sources in the galaxy outskirts (cf. Figure \ref{LF}).
The properties of the closest optical counterpart for each X-ray source are reported in Table \ref{tbl-1}.

Adopting an optical source surface density ranging from 0.05
src/sq.arcsec in the central HST field, to 0.03 src/sq.arcsec in
the southern field, the average chance of a random match with an X-ray
source within our fiducial radius is $\sim\!12\%$, which drops to 4\%
considering only globular cluster candidates which have a factor $3\times$
lower surface density. Assuming a contamination of $\sim\!25$ background
AGNs \citep{Bauer04}, these figures result in NGC~1399 having a fraction
of LMXBs residing in GCs of $f_{\rm GC-LMXB}=65\%\pm 5\%$, in good
agreement with previous estimates \citep{Kim06, Angelini01}. This value,
however, depends on galactocentric distance ranging from $\la\!50\%$
within the central 50\arcsec\ to 68\% (77\%) for $r\!>\!120\arcsec$
($r\!>\!200\arcsec$).

X-ray sources tend to reside in compact bright optical counterparts, i.e.
bright GCs (Figure~\ref{maghisto}): only 28 out of 164 objects ($17\%$) are in fact associated with extended sources. The fraction of
GCs hosting an X-ray source ($f_{XGC}$) drops with magnitude from $\sim\!20\%$ to $2\%$
and depends on radial distance and color of the
host GC (Figure~\ref{LMXB_frac}). 
The drop in the brightest magnitude bins is due to a combination of
the different radial profile of the red and blue GC populations
($\S$\ref{rad_dep}) coupled with the lack of bright red GCs, compared to
the blue population, in particular in the galaxy center
(cf.~Figure~\ref{col_mag})\footnote{The median GC sample color is $(V_{\rm
F606W}\!-\!T1)\simeq0.7$ mag.}. We do not observe any significant
difference in $f_{XGC}$ in the red and blue sub-population
as a function of radius when also splitting in luminosity bins, although
our sample is too small to draw definitive conclusions. In any case we stress that this
interplay between GC galactocentric distance, magnitude, and color must be taken into account
in studies observing only the central regions of galaxies.

Finally note that these results are affected relatively little by our GC selection criteria, since the majority of  X-ray sources matched to an optical counterpart reside in compact objects (see $\S$\ref{struct_parms}): even considering that we may be missing part of the most extended GCs (see $\S$\ref{opt_data_sect}), this would result in a lower $f_{XGC}$ by a few percent and anyway within the statistical uncertainties. 


\subsection{The X-ray Luminosity Function}
For each X-ray source in our catalog, the ACIS Extract procedure (see \S \ref{Chandra_data})
computes the incident photon flux, applying the quantum efficiency 
and spectral response corrections (i.e. using the ARF and RMFs) appropriate for each observation at 
the specific detector location, which are then combined in a final weighted average.
This ensures that the position and time dependence of the ACIS efficiency is properly taken into account.

\begin{figure}[]
\includegraphics[width=8.5cm]{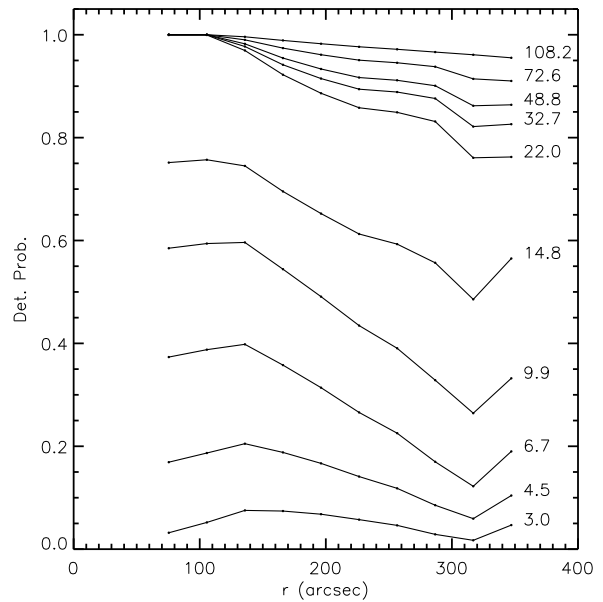}
\caption{X-ray detection probabilities as a function of the off-axis angle and the number of counts detected in a given source. This grid has been calculated following the "forward"
procedure described in \cite{Kim03}. Such probabilities are used to correct for incompleteness (in e.g. LFs, radial profiles) due to the diffuse X-ray background and variable PSF. }
\label{det_prob}
\end{figure}

The X-ray luminosity function (LF) of LMXBs in NGC~1399 was obtained
applying an average conversion factor to the photon fluxes measured by
AE, computed assuming a power-law spectrum with $\Gamma=1.5$ and a Galactic column density of $1.3\times 10^{20}$ cm$^{-2}$, which
corresponds to the spectral model for our average source
(Figure~\ref{hr_col}). 
To correct for contamination due to background
sources we used the AGN number counts of \cite{Bauer04}. The cumulative LF of both GC- and field-LMXBs, shown in Figure~\ref{LF} (left panel), has a
power-law shape down to $\sim 2\times 10^{38}$ erg s$^{-1}$. At fainter
fluxes the combined effect of incompleteness and source variability may
affect the LF shape. To correct for incompleteness, due to the variable
PSF over the FOV and the diffuse X-ray emission, we adopted the ``forward''
procedure described in \cite{Kim03}, accounting for the effect of
background, source counts and distance from the aimpoint. The detection
probabilities as a function of source number counts, at various off-axis
angles, are shown in Figure~\ref{det_prob}.

Since X-ray binaries are intrinsically variable, finite integration
times may influence their detectability. Moreover, the effect of
variability on stacked observations may change the LF slope close to the
completeness limit of the individual observations, as discussed in detail
by \citet{zez07}. We verified that the LF of the individual
\textit{Chandra} observations (\#319 and \#1472) are consistent within the
statistical uncertainties, except for the somewhat shallower completeness
limit due to the shorter exposure time.

The completeness corrected GC-LMXB LF follows a power-law down to $f_X\lesssim
10^{38}$ erg s$^{-1}$ with a differential slope of $-1.7\pm 0.2$ in the $0.8-5\times 10^{38}$ erg/s range,
in good agreement with the one derived by
\cite{Kim06} from a sample of six early-type galaxies. Some residual
incompleteness can be observed in the faintest bins ($\la\!10-20$ counts) mainly due to
the very high background in the galaxy center. In fact splitting the LF in
radial bins (Figure \ref{LF}, right panel) shows that the intermediate bin
(120\arcsec-180\arcsec), which covers the range of maximum completeness
(Figure~\ref{det_prob}) does not deviate significantly from a power-law
down to $10^{38}$ erg s$^{-1}$ and below. The GC-LMXB LF in the outermost radial bin presents an excess of bright ($L_X\!>\!5\! \times\!10^{38}$ erg s$^{-1}$) sources with respect to the inner regions of the galaxy,
suggesting that there may be some residual contamination from background
sources in the galaxy outskirts. In fact only one X-ray source is detected
at $L_X>10^{39}$ erg s$^{-1}$ in the color-selected GC sample (see below).

\begin{figure}[t!]
\centering
\includegraphics[width=8.5cm]{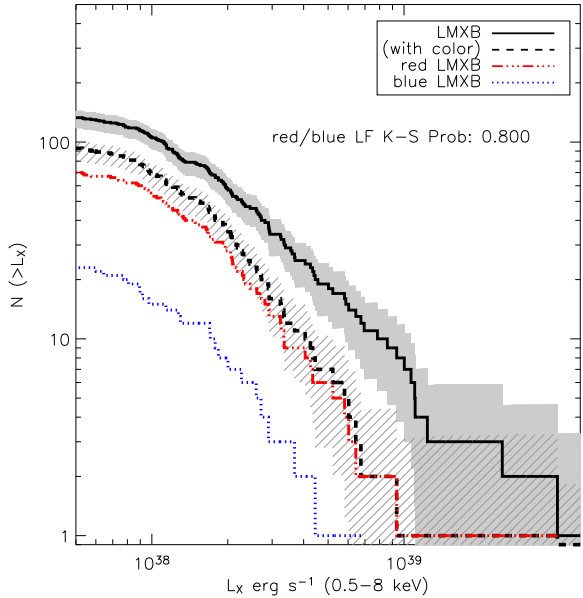}
\caption{Dependence of the cumulative GC-LMXB X-ray luminosity 
function on the color of their host GCs. The K-S probability that red and blue LMXBs are drawn from the same parent population is also shown, indicating that the two LFs are statistically indistinguishable.}
\label{LF_col}
\end{figure}

We do not detect any significant difference in the cumulative LF as a function of the host GC color. Figure~\ref{LF_col} shows that the LFs of red
and blue GC-LMXB are statistically indistinguihable according to a K-S test, and that they are consistent
with the global LF except at the very bright end. 
For $L_X\!>\!10^{39}$ erg s$^{-1}$, two sources included in
the GC-LMXB LF in Figure~\ref{LF} have no color information, while other
three X-ray sources associated with compact optical counterparts have
colors outside the selected range (Table~\ref{col_sel}) and thus likely
represent interlopers.

Our field-LMXB LF is somewhat steeper than the literature estimate, altough consistent within the errors, 
with a differential slope of $-2.5^{+0.7}_{-1.4}$ in the $0.8-5\times 10^{38}$ erg/s range.
Furthermore it suggests a lack of bright LMXBs above $L_X\gtrsim 3\!-\!4\times 10^{38}$:
assuming for field LMXBs the same underlying distribution as observed for GC-LMXBs,
and considering that we have about twice as many LMXB in GC than in the field, the probability
of observing no field source above $5\times 10^{38}$ erg/s is $\sim 1\%$, incresing to $\sim 5\%$
if we allow for cosmic variance in the AGN number counts (which dominate at bright fluxes) by a factor 2.
We find that this significance does not depend on galactocentric
distance, and is also confirmed when analyzing the individual
X-ray observations \#319 and \#1472 separately.

\begin{figure}[]
\centering
\includegraphics[width=8.5cm]{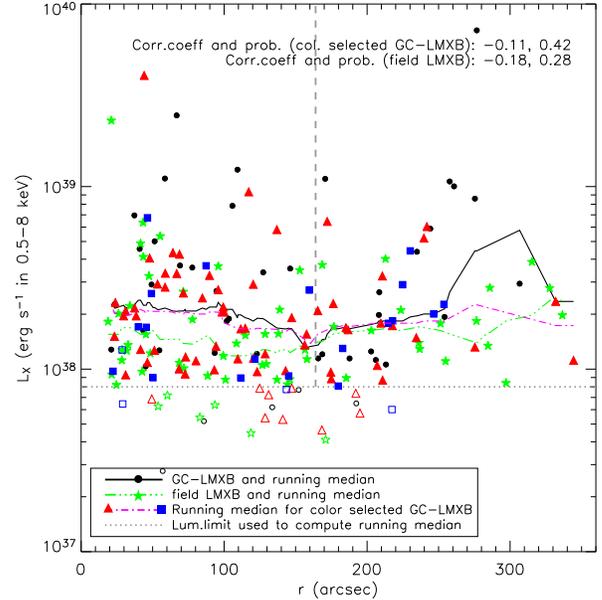}
\caption{LMXB luminosity as a function of galactocentric distance. Triangles and squares represent the red and blue GC-LMXB respectively, while field LMXB are marked by stars. The solid symbols are those used to compute the running average of each subsample, using the uniform completeness limit marked by the dotted line. 
We find that in the galaxy center GC-LMXB are significantly brighter than field LMXBs, although we don't detect the correlation with galactocentric radius reported by \citet[ see discussion in text]{Fabbiano10} for NGC 4278. The vertical dashed line marks the distance limit adopted for the correlation tests, corresponding to the distance limit probed by \citet{Fabbiano10}. }
\label{Lx_r}
\end{figure}
 
The LMXB X-ray luminosity is shown in Figure \ref{Lx_r} as a function
of galactocentric radius. There is some evidence that the median GC-LMXB 
X-ray luminosity is larger than for field LMXBs, at least in the galaxy center
within $r<160$ arcsec, in agreement with the flatter LF observed for GC-LMXBs.
At larger radii the difference disappears but here the
smaller number of sources and the background contamination, which
affects mainly field LMXBs, makes any conclusion tentative.
To check that this difference is not the result of a
sampling effect due to the fact that we are observing about twice as many GC- than field-LMXBs,
we performed 1000 simulations, randomly resampling our LMXB sample, 
while preserving the relative ratio of the two populations. We find that the likelihood
to obtain the median difference that we observed is $\sim 2\%$.

This result is in agreement with what is observed by \cite{Fabbiano10} in NGC 4278:
in that galaxy the authors find an anti-correlation between the GC-LMXB X-ray luminosity and 
galactocentric distance, in the sense that GC-LMXB are brighter closer to the galaxy center, and brighter on average than field sources. In NGC1399 however a Spearman Rank test, within $160$ arcsec\footnote{This radius was chosen rescaling the galactocentric radius of $120''$ probed in NGC 4278 by galaxy distance and $D_{25}$.}, does not yield any significant correlation (see Figure \ref{Lx_r}) between X-ray luminosity and galactocentric distance, as can be expected given the large scatter in $L_X$.

\subsection{X-ray Spectral Properties and Variability}

\begin{figure}[t!]
\includegraphics[width=8.7cm]{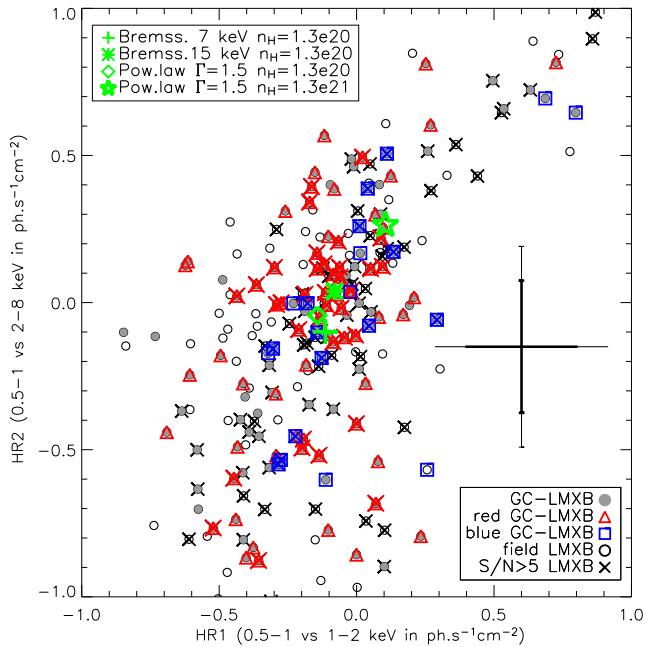}
\caption{HR1-HR2 X-ray color-color diagram for field- and GC-LMXBs. Red and blue solid circles represent LMXBs hosted by red and blue GCs respectively, while black symbols are GC-LMXB without color informations. The crosses mark sources with high-quality data ($S/N\!>\!5$). We overplot the predictions from different spectral models and variable absorption: a Bremsstrahlung with $kT=7-15$ keV, or a power-law model with $\Gamma=1.5$ and column density $n_H=1.3\times 10^{20}$ are all consistent with the bulk of the population, while a higher column density of $n_H=1.3\times 10^{21}$ produces a shift of $\sim 0.3$ toward the upper-right corner of the plot. 
The typical error for all (thin line) and high-quality (thick line) data is shown on the right.}
\label{hr_col}
\end{figure}

We test whether we can detect any dependence of the X-ray spectrum of
LMXBs on the presence of a host GC or on the color of the host GC itself.
For this we compare the $0.5-1$ vs $1-2$ keV (HR1) to
the $0.5-1$ vs $2-8$ keV (HR2) hardness ratios, finding that field-LMXBs,
as well as red and blue GC-LMXBs, 
span a similar range of X-ray colors, consistent with either a {\it
Bremsstrahlung} or power-law spectral model (see Figure~\ref{hr_col}).
A 2D K-S test confirmed that there is no significant difference in the 
hardness ratio of these LMXB populations.

We investigate the time variability of X-ray sources both
within each observation through Kolmogorov-Smirnov testing, and across
observations, comparing the error-weighted fluxes at the two epochs. 
We detect variability in 30 X-ray sources (see Table \ref{tbl-1}),
13 of which residing in color selected GCs. The fraction of
variable sources increases from $\sim\!10\%$ below $3\!\times\!10^{38}\, \mbox{erg s}^{-1}$
up to $\sim\!33\%$ for $3\!\times\!10^{38}\!<\!L_X\!<\!10^{39}\, \mbox{erg
s}^{-1}$, as expected by the better photon statistics, while it remains constant 
at $\sim\!14\%$ as a function of the host GC optical magnitude.
While these results suggest that bright X-ray sources in GCs are not simply due to 
the superposition of several low luminosity binaries, we cannot exclude, based only 
on temporal analysis, that some GC host multiple LMXBs, since the observed variability can be
easily accounted for if, e.g., a bright source dominates the LMXB population within a GC.

On the other hand in the only color-confirmed GC-LMXBs (source no.141 in Table \ref{tbl-1})
with $L_X> 10^{39} \mbox{erg s}^{-1}$, we do not detect
any sign of variability despite the $S/N>20$. To test the statistical 
significance of this result we extracted from the RXTE archive\footnote{\url{http://heasarc.gsfc.nasa.gov/docs/xte/recipes/mllc\_start.html}} 
the Mission-Long lightcurve 
of the Galactic BH binary GRS1915+105. After degrading the data to the same S/N level of
our source, we find that the likelihood of finding a difference of less than 3\% in flux 
between 2 observations obtained 3.3 yrs apart, as for NGC1399, is only $\sim 3\%$.



\begin{figure*}[t!]
\includegraphics[width=8.5cm]{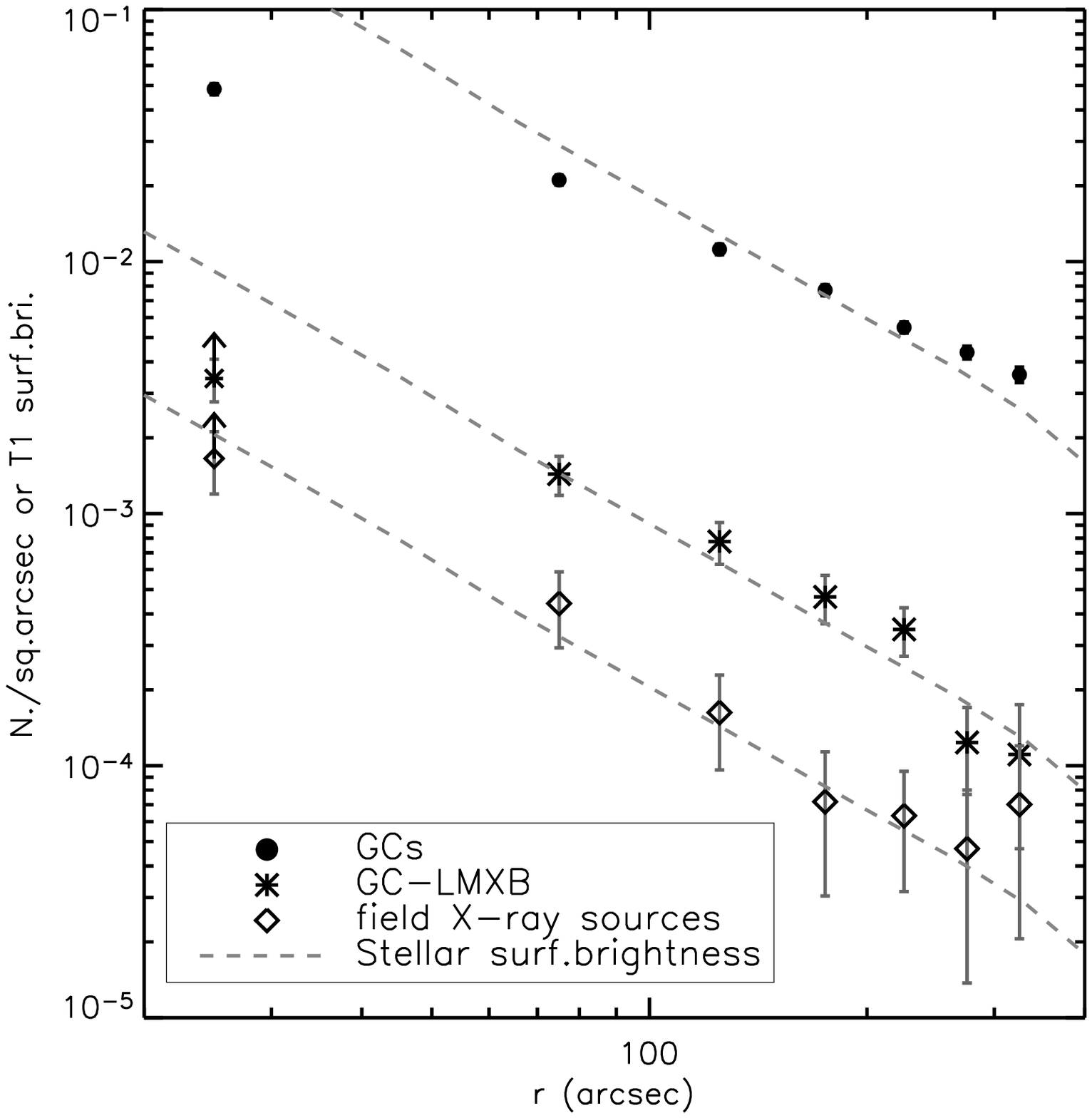}
\includegraphics[width=8.5cm]{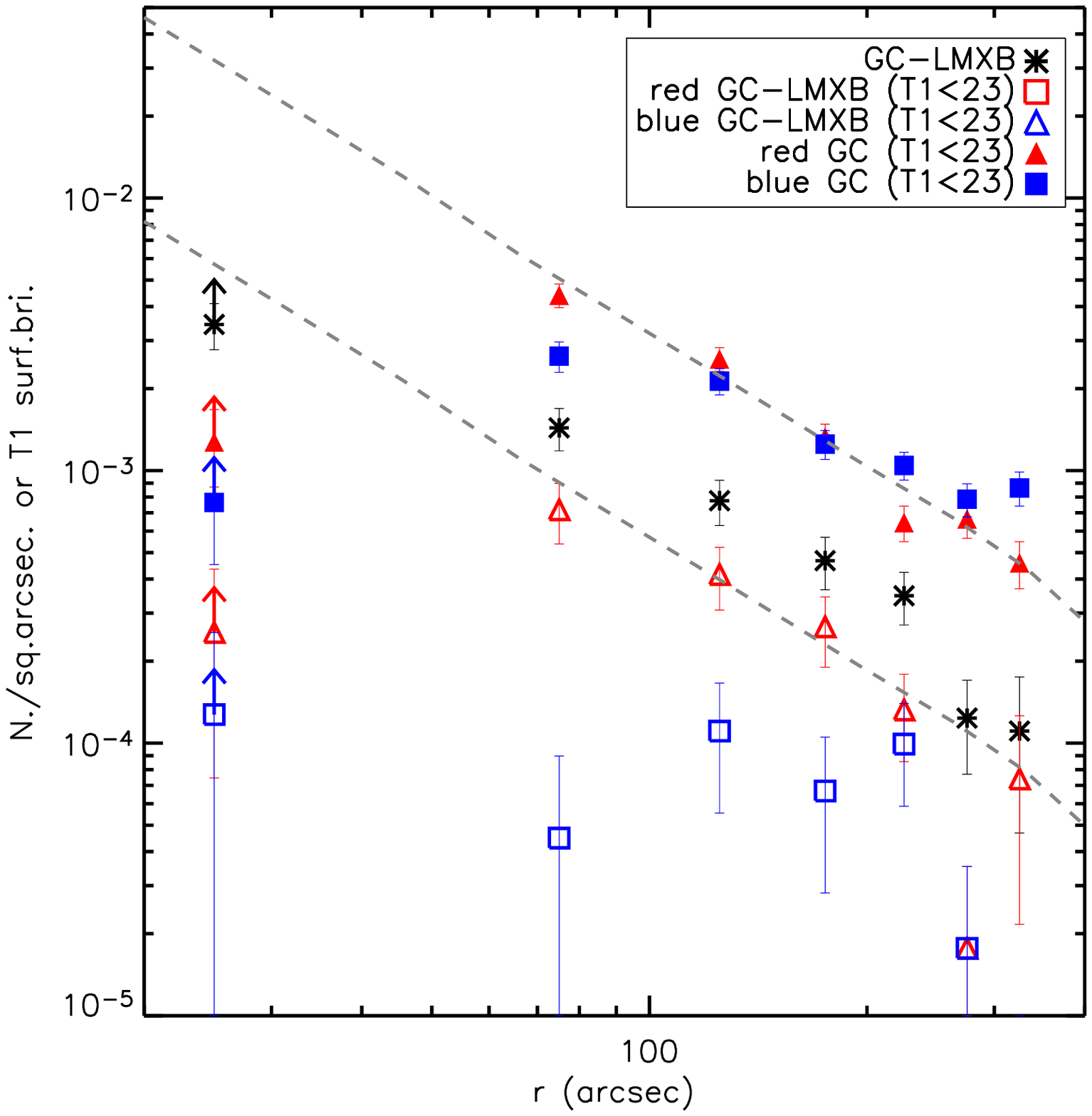}
\caption{Radial surface density profiles of GCs and LMXBs. The dashed lines represent the T1 surface brightness profile of the diffuse galaxy light, taken from \cite{Dirsch03} and arbitrarily rescaled. GC-LMXBs tend to follow the parent GC population, while field LMXBs are closer to the galaxy light distribution.}
\label{rad_prof}
\end{figure*}

\subsection{LMXB/GC connection: Spatial distribution}
\label{rad_dep}

As already discussed in $\S$\ref{matching} the fraction of LMXBs hosted
in GCs changes with galactocentric radius, indicating different
spatial density distributions of field- and GC-LMXBs. In
Figure~\ref{rad_prof} we plot the radial profiles of the optically
identified GC population, the GC-LMXBs and field X-ray sources.
Incompleteness effects in the LMXB profiles have been corrected as
described in previous sections, except for the central bin which
represents a lower limit since the high level of diffuse emission makes the
correction uncertain. As indicated by other studies
\citep[e.g.][]{Dirsch03, puzia04, Bassino06, Schuberth10} the GC
population is more extended than the galaxy light. A similar behavior is
found for LMXBs hosted in GCs (even though with less significance due to
the smaller sample statistics) while the field X-ray sources have a
steeper profile close to the one of the diffuse galaxy light. Note that
the surface density of GCs within the central 50\arcsec\ is lower than
expected from a simple power-law extrapolation of the external GC
distribution. This is not an incompleteness effect since it is present even
when only the brightest GCs are taken into account, and is in agreement
with the shallower central profile observed by \cite{Dirsch03},
\cite{Bassino06}, and \cite{Schuberth10}. Thus, while the
difference between GC and field LMXBs is mostly due to the central 
region of the galaxy ($r<50``$) where the completeness corrections are 
more uncertain, the fact that the GC-LMXB surface density profile 
presents a similar deficit of sources in the
central bin, while field-LMXBs show no such behaviour, 
supports the view that this difference is not due to incompleteness effects.

In Figure~\ref{rad_prof} (right panel), we further divide the GC
population according its $C\!-\!T1$ color (see Table \ref{col_sel}). In this case the
incompleteness of the color catalog is clearly visible within the
central bin, as discussed in $\S\ref{opt_data_sect}$.~However,
the plot shows that the shallower GC distribution is mainly due to the
blue GC component \citep[see also][]{Schuberth10}. When compared with
GC-LMXBs we find no significant difference between the LMXB distribution
and the one of the host GC population.

\begin{figure}
\includegraphics[width=8cm]{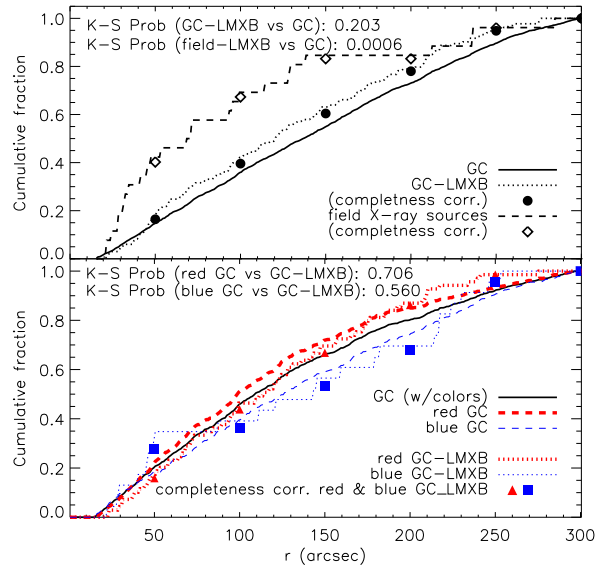}
\caption{{\it Upper panel:} Cumulative surface density profiles of GCs and LMXBs. The low K-S test probability  reported in the figure confirms that field LMXBs have a different radial distribution from the GC population.   {\it Lower panel:} Cumulative profiles of GCs and LMXBs with available color informations. Both red and blue GC-LMXBs are distributed according to their host GC population. In both panels the points show that incompleteness effects do not affect the results.}
\label{cum_prof}
\end{figure}

These results are confirmed by the cumulative distributions in
Figure~\ref{cum_prof}: while the GC and GC-LMXB profiles are consistent
with each other (upper panel), the field-LMXBs are much more concentrated
than GCs at $>99.9\%$ confidence level according to a K-S test. Furthermore, the blue GC
sub-population is more extended than the red one \citep{Dirsch03}.
However, LMXBs seem to follow the distribution of their host GCs, with 
little evidence of the more extended distribution found by \cite{Kim06}. 
Figure~\ref{cum_prof} also shows that incompleteness effects in the X-ray source distribution do not
significantly affect these conclusions.

\begin{figure}
\includegraphics[width=8.5cm]{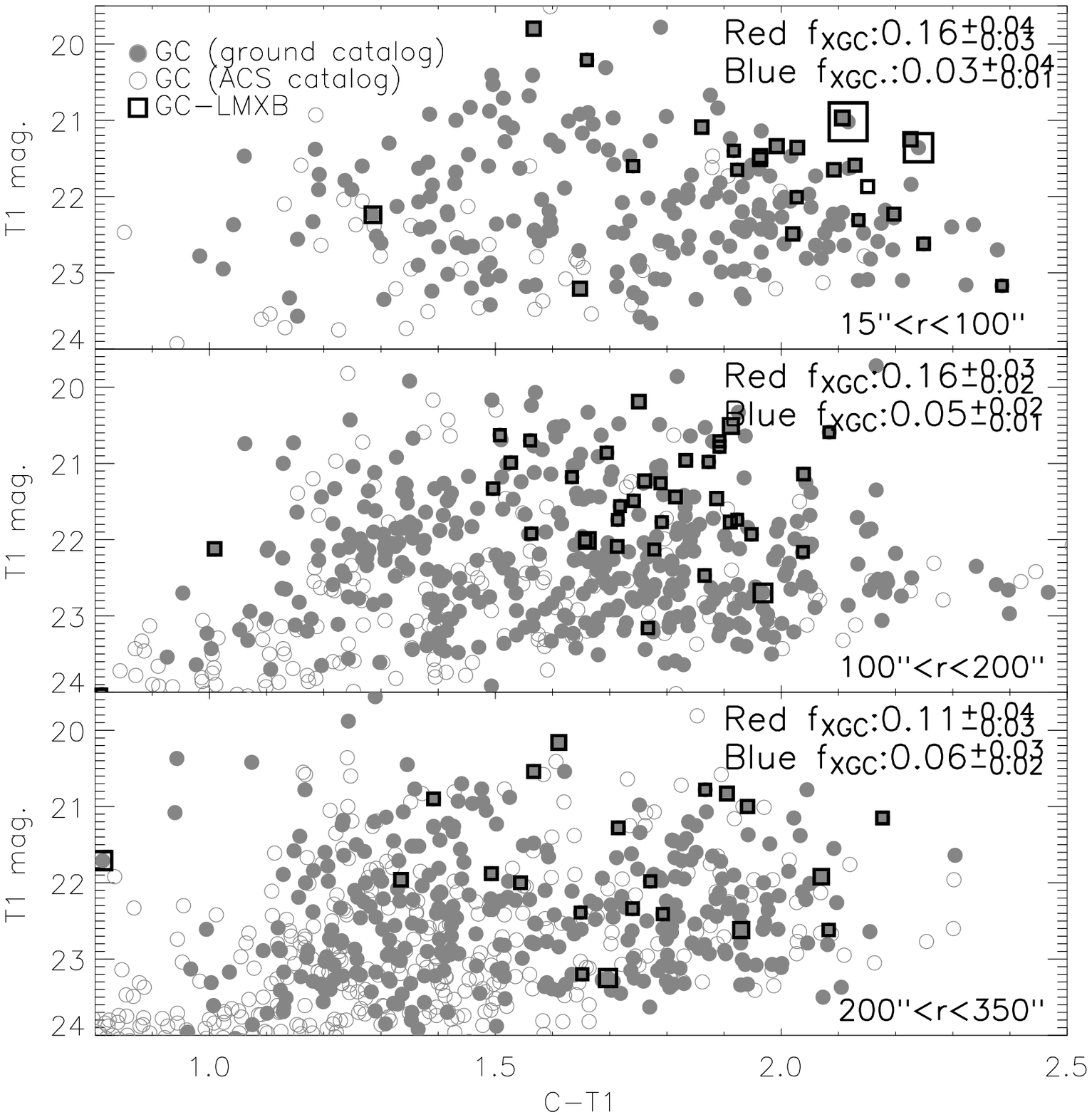}
\includegraphics[width=8.5cm, bb=54 144 558 350]{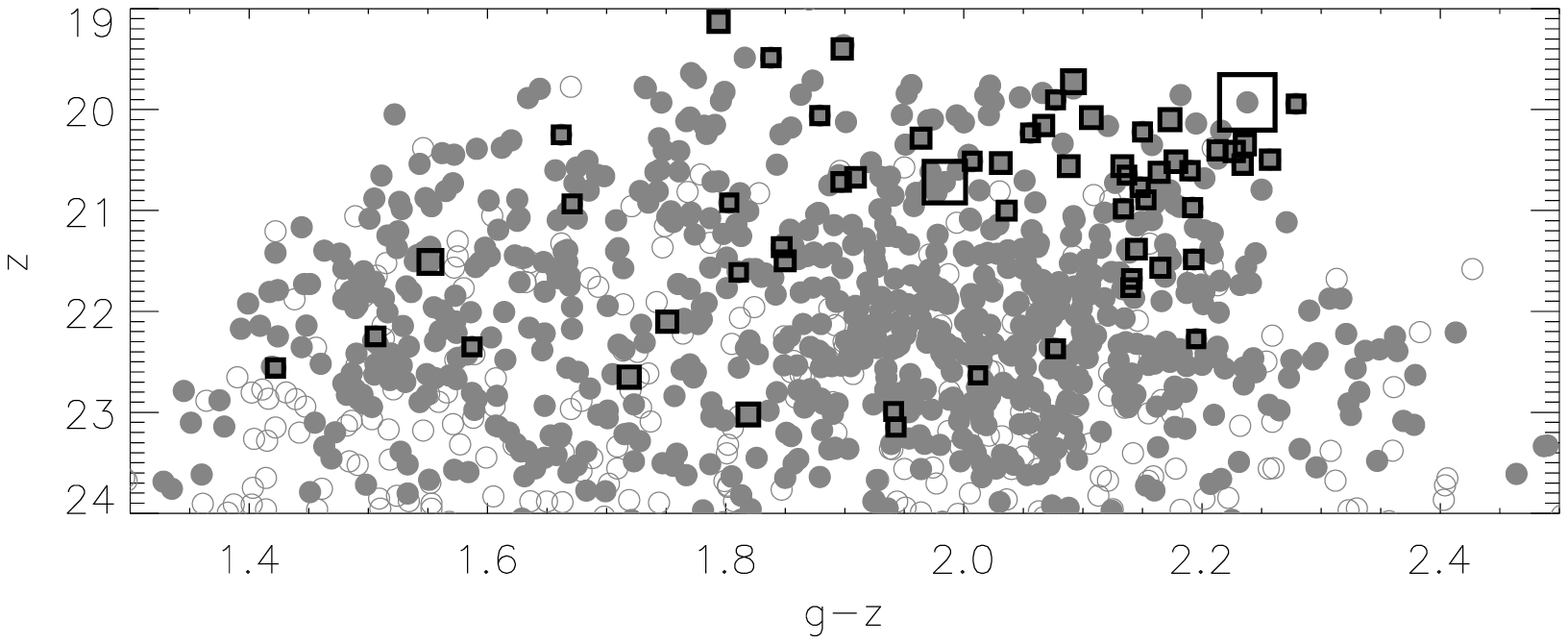}
\caption{Color-magnitude diagrams using $C\!-\!T1$ ground-based (upper panel) and $g\!-\!z$ HST photometry (lower panel). Ground-based plots are split according to galactocentric distance, while HST data refers to the whole ACS field ($\sim\!200\arcsec\!\times\!200\arcsec$). Open grey dots represent sources in color catalogs while solid ones refer to GC candidates detected in our F606W ACS data. Black squares indicate GC hosting LMXBs, where the symbol size proportional to the $\log$ of the X-ray flux in the $0.5\!-\!8$ keV band. The fraction $f_{XGC}$ of 
color-selected GCs hosting X-ray sources, reported in the upper panel, shows no dependence on galactocentric distance.}
\label{col_mag}
\end{figure}

This behaviour is also seen in the color-magnitude diagrams shown in
Figure \ref{col_mag}: not only do the LMXB-host GCs become bluer toward
the galaxy outskirts, but the overall GC population shifts toward bluer
colors. In addition, we see the presence of a very red GC population
($C-T1\!\approx\! 2.15$ mag or $g-z\!\approx\!2.2$ mag), already noticed
by \cite{Kundu07}, which resides in the NGC~1399 core and hosts the
majority of red LMXBs.
The fraction of red and blue GCs hosting LMXBs is shown in the plots. We
can see that both red and blue GCs have a constant frequency within the
errors, as expected if LMXBs follow the distribution of their host GCs.
While the mild decrease observed in the red population in the outermost
bin is marginally consistent with the trend reported by \cite{Kim06}, the
LMXB frequency within the blue population shows the opposite trend,
inconsistent with the large drop ($\gtrsim\!2$) predicted by these authors.

\begin{figure*}[t]
\centering
\includegraphics[width=12.5cm, bb=54 144 558 648]{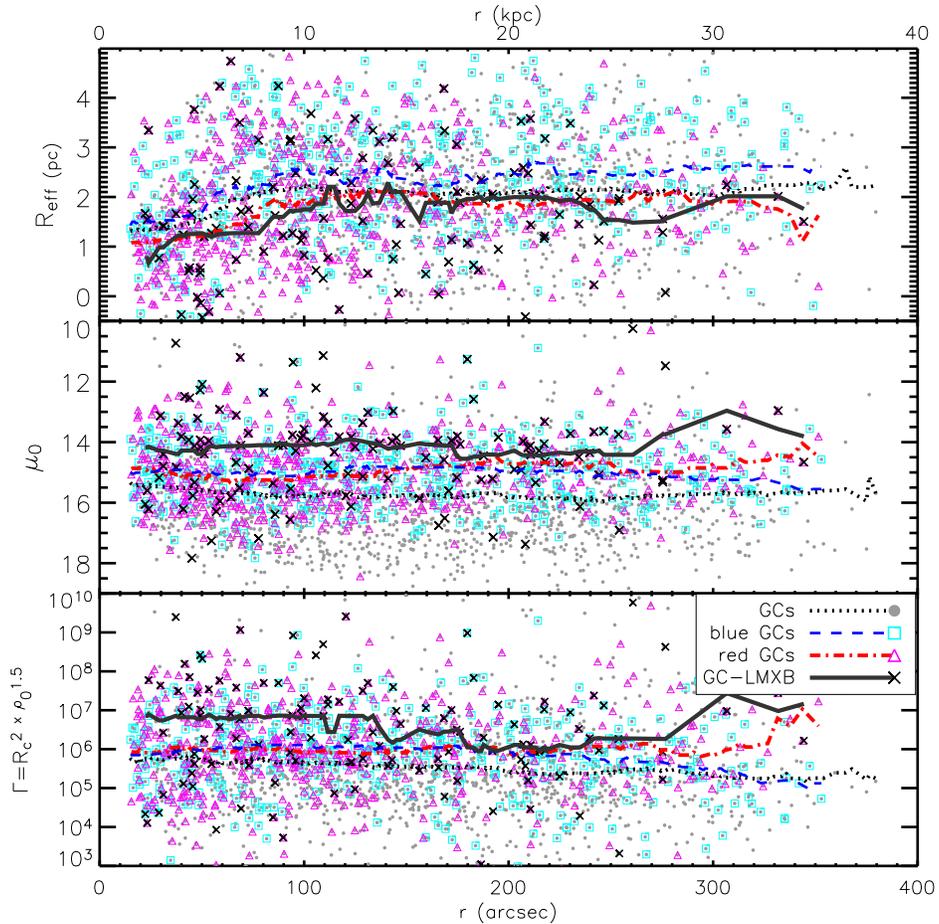}
\caption{Galactocentric dependence of GCs structural parameters. GCs and color-selected sub-populations are shown in grey dots, red triangles and blue squares, respectively. GCs hosting LMXBs are marked by black $\times$. The running median values (solid and dashed lines) show that the different populations have different structural parameters.}
\label{rad_size}
\end{figure*}

\subsection{The LMXB/GC Connection: GCs Structural Parameters}
\label{struct_parms}

Structural parameters were measured, as explained in detail in P11, using the {\sc
GalFit}\footnote{http://users.obs.carnegiescience.edu/peng/work/galfit/gal
fit.html} software \citep{Peng02} to fit a \citet{King} model to our HST data,
including the error maps produced by the {\sc Multidrizzle} routine as weight
maps, we derived tidal, core and effective radii, and central surface
brightness values. 

Typical GCs are marginally resolved at the distance of NGC~1399, hence
accurate knowledge of the PSF is crucial to derive robust GC structural
parameters. As discussed in detail in P11, we designed a specific
software, the {\sc Multiking} package\footnote{The {\sc Multiking} package
and documentation is available at
\url{http://www.na.infn.it/$\sim$paolillo/Software.html}} which makes use
of the empirical PSF library for ACS/WFC \citep{anderson05, anderson06} to
build a new drizzled PSF library replicating the actual data frame
properties (orientation, dither pattern, astrometry, etc.) in order to
account for all effects related to the observing strategy and data
reduction process.

The accuracy of our measurements was estimated using several thousand simulated GCs produced with the {\sc Multiking} package to include all
instrumental effects (field distortions, PSF variation, etc.).  We used the simulated GC catalog to correct 
for residual systematics affecting the structural parameters measurements, by fitting a polynomial
function to the measured values as a function of the input value, and applying such correction to
the real GC measurements. While we refer the reader to P11 for a
comprehensive discussion of the fitting accuracy, our simulations show that we can robustly measure the individual effective (i.e.~half-light) GC radius in the range $\sim\!1.5\!-\!20$ pc with an average
uncertainty of $0.56$ pc and little dependence on magnitude, background
level or galactocentric distance.
We also emphasize that our bona-fide GC sample has an integrated $S/N>100$, in good agreement with the minimum prescription 
of \citet{Car01} to measure sizes of marginally resolved GCs\footnote{\citet{Car01} claim that $S/N>500$ is required in order to fully recover all King model parameters for every type of GC out to a distance of $\sim 40$ Mpc, i.e. twice as far as NGC 1399. On the other hand they state that $S/N\gtrsim 100$ is appropriate for, e.g. Virgo galaxies,
or less concentrated systems, and that half-light radii are recovered with even better accuracy. Furthermore our spatial sampling (pixel size) is $\sim 3$ times better than what was used in their study.}.

In the top panel of Figure \ref{rad_size} we show the galactocentric dependence of the half-light radius. 
In the central 50\arcsec\ the distribution is dominated by a very
compact GC population, which also hosts the majority of LMXBs. An
increase of the GC effective radius with galactocentric distance has
originally been observed in the Milky Way \citep{VdB91} and later in several other massive early-type galaxies \citep[e.g.][
and references therein]{Spitler06, Madrid09, Harris09}. In NGC~1399 the 
projected size gradient seems to be mostly
confined to the inner core ($\la\!100\arcsec$), and remains
approximately constant outside $\gtrsim\!1\, R_{\rm eff}$ out to
$\sim\!5\, R_{\rm eff}$ of the diffuse galaxy light, similar to what was found by 
\cite{Spitler06} in the Sombrero galaxy. However, at odds with the latter work, we find the same behaviour for both red and blue subpopulations even at large radii, arguing in favor of an \textit{intrinsic} difference and against projection effects (see also $\S$\ref{Disc:struct}).
This may suggest the presence of a lower threshold in radius below which tidal forces allow the survival of
most compact stellar systems and further supports the need to reach larger galactocentric distances 
than probed by most high spatial-resolution HST studies of extragalactic GCs, since the very central regions
of giant galaxies do not necessarily reflect the properties of the entire GC population.

\begin{figure}[ht!]
\includegraphics[width=8.5cm]{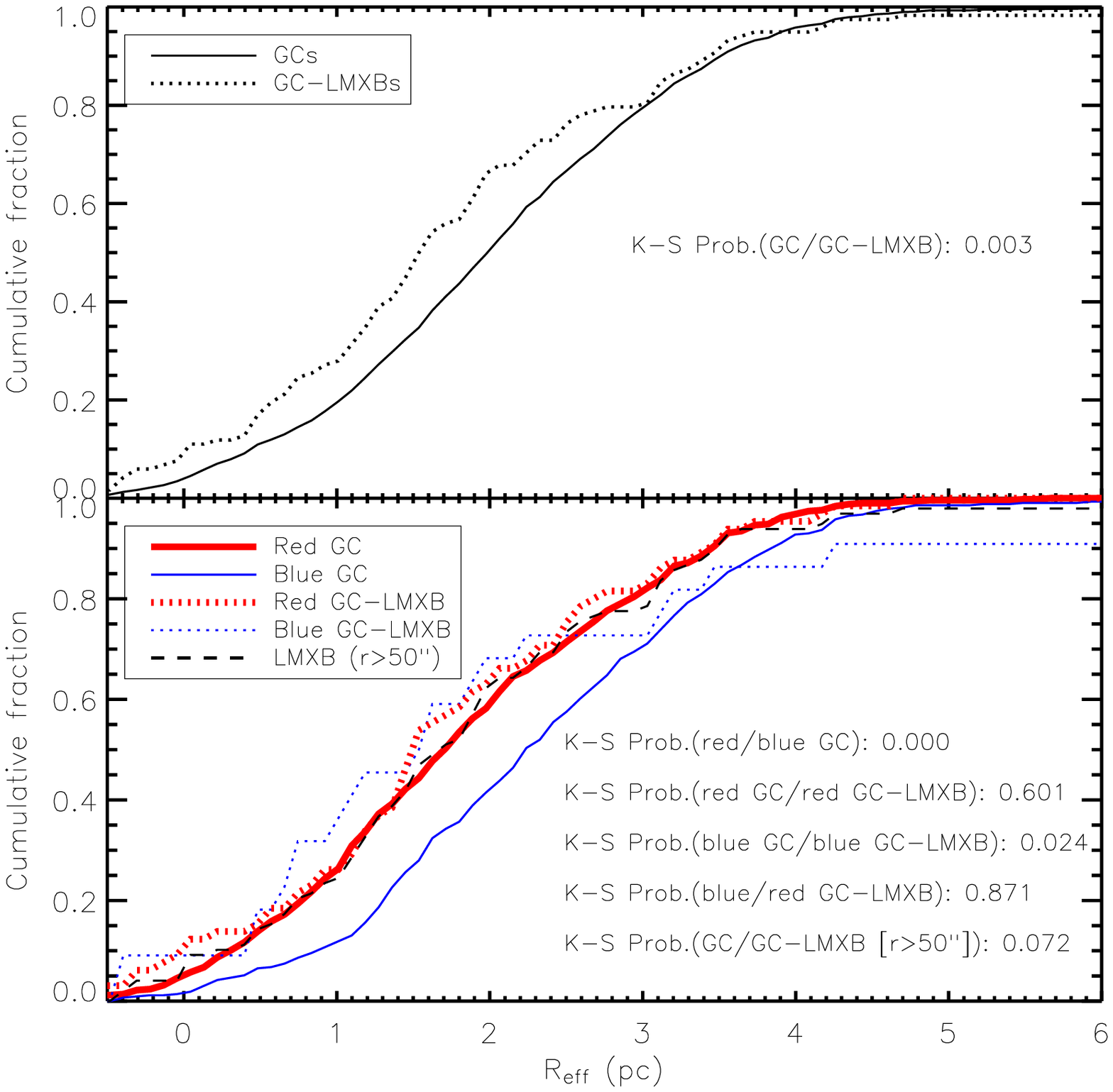}
\caption{Cumulative distributions of GC effective radii. The upper panel illustrates the distributions for GCs with and without LMXBs, while the lower panel shows the corresponding distribution for the blue and red GC sub-populations, as well as all LMXB sources. The K-S test results, reported in the plot, show the probability of different subsamples to be drawn from the same parent distribution: GC-LMXBs are more compact than the bulk of the GC population, but are similar to the red GC subpopulation.}
\label{reff_cum}
\end{figure}

We find that GCs hosting X-ray sources (XGC) are, on average, more compact that the rest of the GC population, but
do not differ significantly form the red GC population which hosts the
majority of LMXBs. This is clear from looking at the cumulative
distribution of the GC effective radius shown in Figure~\ref{reff_cum}:
while red and blue GCs have different sizes at a significance level
$>\!99\%$, we cannot detect any significant difference between red
XGCs and the overall red GC sub-population. On the other hand, LMXBs residing in blue GCs seem to prefer the most compact systems, even though this difference is only significant at the $2\sigma$ level. This indicates
that more than one physical parameter is driving LMXB formation.

\begin{figure*}[t]
\centering
\includegraphics[width=10cm, bb=60 144 558 630]{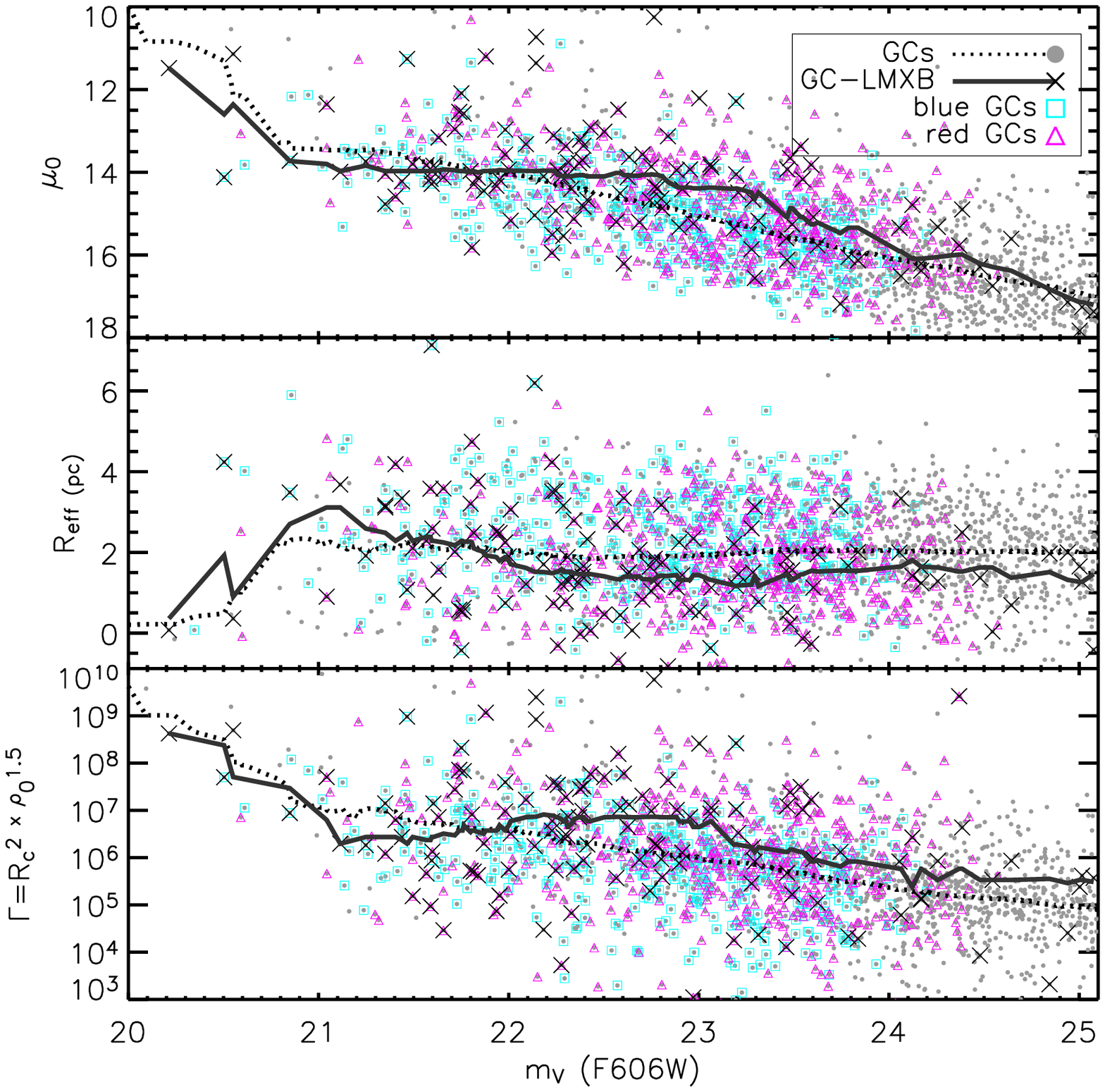}
\caption{Structural parameters versus GC magnitude. Solid circles represent GCs, while crosses indicate LMXBs. The solid and dashed lines represent the running median of the GC-LMXB and GC distribution, respectively. LMXBs tend to have different structural parameters from the GC population, except at bright magnitudes.}
\label{size_mag}
\end{figure*}

\begin{figure}[]
\includegraphics[width=8.5cm]{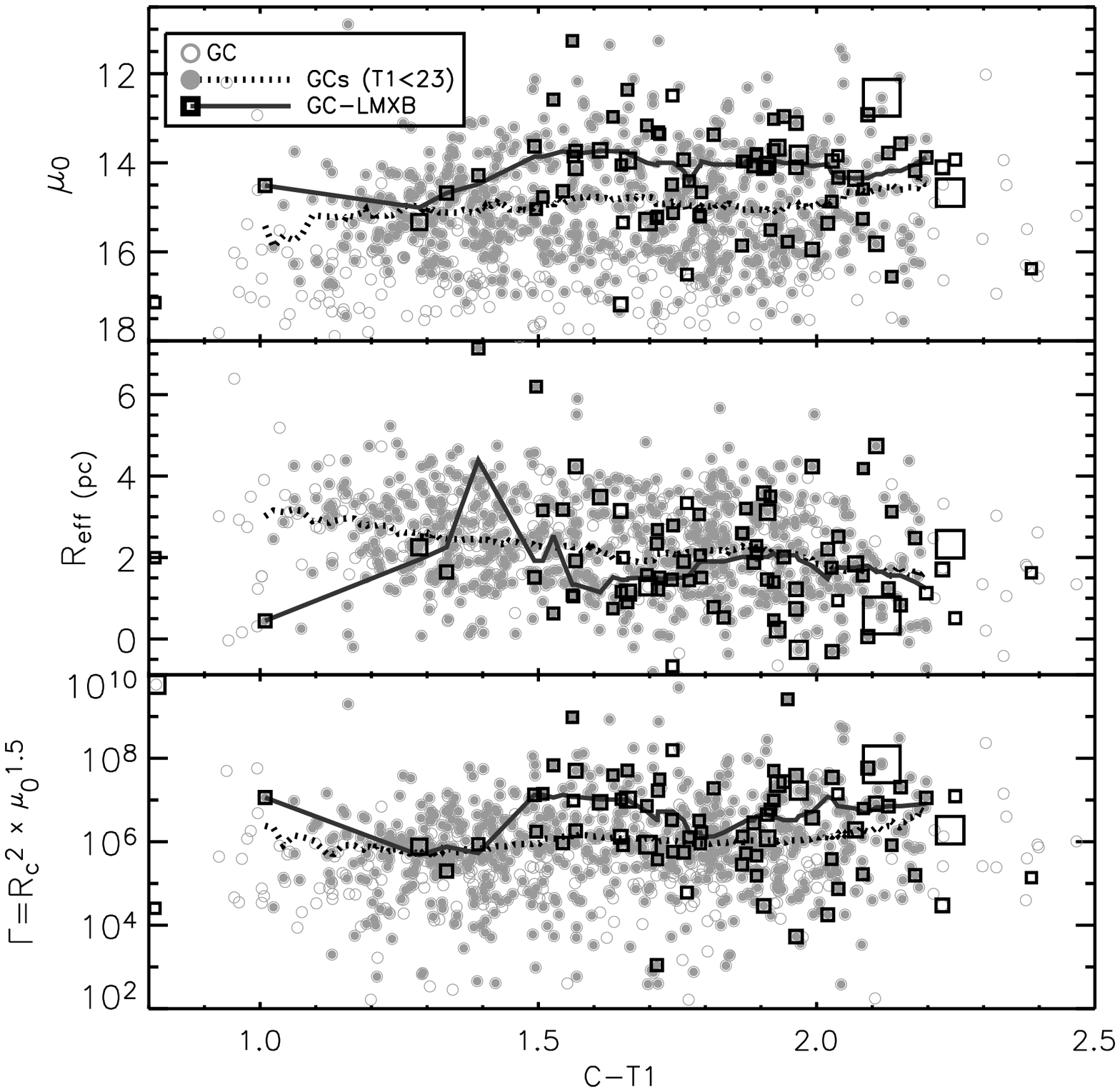}
\caption{Structural parameters versus GC color. Grey solid (open) circles represent GCs above (below) the magnitude thereshold of T1=24 used for the color-selected subsample of Table \ref{col_sel}, while LMXBs are marked by black squares whose size is proportional to the logarithm of their X-ray luminosities. The solid and dashed lines represent the running median of the GC-LMXB and GC distribution, respectively.}
\label{size_col}
\end{figure}

The middle and bottom panels of Figure~\ref{rad_size} show the $V_{\rm F606W}$-band
central surface brightness $\mu_0$, and the interaction rate $\Gamma = R_{\rm c}^2 \rho_0^{1.5}$
\citep[see e.g.][]{Verbunt06}, as a function of galactocentric radius. The Figure indicates that
LMXBs are preferentially hosted by high central surface brightness and/or
high interaction rate GCs at all galactocentric radii, independent of the
host GC color. This behaviour, however, is due in part to the tendency of
LMXBs to reside in bright GCs discussed in \S\ref{matching}. To understand whether GC
structural parameters have an effect on LMXB formation likelihood in
addition to luminosity and metallicity, we plot the main parameters
against GC luminosity in Figure~\ref{size_mag} and against GC color in
Figure~\ref{size_col}. While it is difficult to disentangle the effect of 
each parameter since they are correlated through the definition of the
King profile, LMXBs residing in intermediate-luminosity and faint GCs ($m_V\!>\!22$)
have a higher central surface brightness, smaller effective radius and larger
encounter rate than the average GC population. However, at variance with
other studies \cite[e.g.][]{Peacock09} this difference disappears for the
brightest GCs.

Figure~\ref{size_col} shows that GC color has an intrinsic effect on the
LMXB formation likelihood, since blue GCs appear deficient in LMXBs even
though they have similar central surface brightness 
and interaction rates as the red ones, except perhaps for very blue
colors ($C\!-\!T1\!<\!1.3$ mag). We conclude that the different fraction
of LMXBs observed between the red and blue GC sub-population cannot be
attributed to  differences only in their structural parameters, thus,
supporting the view that stellar evolution \citep{Grindlay87, MKZ04,
Ivanova06} and/or mass segregation effects \citep{Jordan04} must 
influence LMXB formation.

To quantify the likelihood that GC structure has an impact on LMXB
formation, in addition to luminosity and color, we resampled the photometric GC
dataset to match the XGC optical luminosity and $C-T1$ color distribution. We generated 10000
resampled distributions testing through S- and T-statistics whether the
mean and variance of the resampled dataset is equivalent to the
reference XGC distribution. Table~\ref{resamp_tab} summarizes the
probabilities of obtaining the same median for the LMXB and non-LMXB 
GC population for a given GC parameter. This Monte-Carlo exercise shows 
that XGCs tend to have significantly ($P\gtrsim 99\%$) larger $\Gamma$
and smaller $R_{\rm eff}$ than the
parent GC population even after removing the luminosity (mass) and color dependence.
Similarly, by resampling according to the XGC distribution in either
$\mu_0$ or $\Gamma$ and color, we find that GCs hosting LMXBs are
more likely to be brighter than the rest of the GC population,
indicating that the GC mass seems to play a role even after removing the
dependence on the other GC parameters. This is partially at odds
with the findings of \citet{Jordan07a} in the Cen-A GC system, where mass
does not seem to play a significant role in LMXB formation, but can be explained
from the comparison shown in Figure \ref{size_mag} where
structural parameters of XGC are significantly different from the whole GC population
only at faint ($m_V\gtrsim 22$) magnitudes while at the bright end this difference disappears. 

\begin{deluxetable}{ccccc}[t!]
\centering
\tabletypesize{\scriptsize}
\tablecaption{Two-parameter resampling probabilities\label{resamp_tab}}
\tablehead{\colhead{Resampled in} & \colhead{P($m_V$)} & \colhead{P($\mu_0$)} & \colhead{P($\Gamma$)} & \colhead{P($R_{\rm eff}$)} } 
\startdata
$m_V$, C--T1 color & \nodata & 0.26 & 0.001 & 0.01\\ 
$\mu_0$, C--T1 color & 0.003 & \nodata & 0.16 & 0.22\\ 
$\Gamma$, C--T1 color & 0.002 & 0.04 & \nodata & 0.39\\ 
\enddata
\tablecomments{The values represent the probability of obtaining the same 
median for the LMXB and non-LMXB GC populations, after resampling the 
non-LMXB sample according to the probability density distribution of the
two quoted parameters. See text for details.}
\end{deluxetable}

\section{Discussion}
\label{discussion}
Our analysis confirms the result, first reported by \citet[][ see also
\citealt{Kim04}]{Angelini01}, that NGC~1399 has the highest global fraction of
LMXBs hosted by GCs ($f_{\rm GC-LMXB}$) among early-type galaxies studied so far, with $f_{\rm GC-LMXB}=65\%\pm 5\%$. We also find however that $f_{\rm GC-LMXB}$ increases with galactocentric distance, ranging from
$\sim 50\%$ within 50\arcsec\ ($\sim 5$ kpc) to $>75\%$ at $r>200\arcsec$ ($\gtrsim 20$ kpc). 
While the exact value of $f_{\rm GC-LMXB}$
is affected by large uncertainties, in particular near the center of the
galaxy due to the high X-ray background, an increase of $f_{\rm GC-LMXB}$ 
with galactocentric distance is in agreement with
the fact that the radial distribution of GCs in NGC~1399 is more extended
than the galaxy light, while field LMXBs tend to follow instead the galaxy
surface brightness profile ($\S$ \ref{rad_dep}). Thus, some care must be
taken when comparing different $f_{\rm GC-LMXB}$ estimates, sampling
different galactocentric distances. In this respect we note that our
result suggests a lower central $f_{\rm GC-LMXB}$ than the one measured by \citet[][ $f_{\rm GC-LMXB}\sim 70\%$]{Angelini01}. 

While $f_{\rm GC-LMXB}$ is known to depend on GC specific frequency
$S_N$ \citep{MKZ03, Sarazin03, Juett05}, NGC~1399 has a large
$f_{\rm GC-LMXB}$ even after taking into account its high $S_N$. Our
results suggest that this galaxy is only marginally consistent (at the
$3\sigma$ level) with the $\sim 50\%$ value found for galaxies with
similar $S_N$ by \cite{Kim06}, such as NGC 4649 or NGC 4472 (see their
Figure~15). This difference is not affected by the intrinsic radial gradient
in the GC-LMXB distribution, since the \cite{Kim06} study used wide-field
ground-based data covering a galaxy fraction similar to our work; furthermore, if the
\cite{Kim06} GC sample is contaminated by background galaxies (hosting an X-ray source),
the actual $f_{\rm GC-LMXB}$ will be lower than observed by these authors thus strengthening
our conclusion. Its significance, however, also depends on the 
probed X-ray luminosity range, as discussed further in $\S$\ref{Sec:spatial_distro}.

On the other hand, on average $\sim\!6\%$ of GCs with $m_V\!<\!26$
($M_V\!<\!-5.5$) host LMXBs; while this number is consistent 
with the average value for early-type systems reported in literature
\cite[see][ and references therein]{Fabbiano06}, we point out that
the exact value depends on the studied magnitude range
(Figure~\ref{LMXB_frac}) and, to a lesser extent, on galactocentric
distance (Figure~\ref{col_mag}). In particular our data allow us to probe the
GC population $\sim 2$ magnitudes below the LF turnover, while most 
studies based on color selected samples are limited to the bright LF end (Fig. \ref{maghisto}). 
LMXBs are preferentially hosted by bright ($m_V\lesssim 24$) and red GC
with fractions $f_{XGC}$ that can be $>20\%$ (Figure \ref{maghisto} and
\ref{LMXB_frac}). This value is almost twice as large
as the one found in similar early-type galaxies by, e.g., \cite{Kim06}; 
the depth of the X-ray data can only account for part 
of this difference since \textbf{i)} our completeness limit
is comparable to the one of the combined LF of the \cite{Kim06} 
sample and \textbf{ii)} we have $<30\%$ more X-ray sources in the 
merged dataset than in the \#319 observation 
used by the latter authors for NGC1399 (see $\S$ \ref{Chandra_data}).
On the other hand the observed difference could be explained if 
the ground-based data are significantly contaminated by background 
sources, which would lower $f_{XGC}$ measured by \cite{Kim06}.

Red GCs are $\gtrsim 3$ times more likely to host an
LMXB than blue GCs, in agreement with the results of \cite{Kundu07}.
We confirm the presence of a very red GC population which hosts most
LMXBs in the galaxy center (Figure~\ref{col_mag}), as reported by
\cite{Kundu07}, and which is better visible using the larger $g\!-\!z$
color baseline of the ACS dataset. However this population seems to
disappear at larger galactocentric distances, where the overall GC
population, including those hosting LMXBs, moves toward bluer colors.
While the presence of a similar very red sub-population was not observed
in the other 4 galaxies studied by the latter authors, a more
homogeneous dataset probing large galactocentric distances, is needed to
address the problem of the universality of such feature and its role in
LMXB formation.

\subsection{Insights from the LMXB Spatial Distribution}
\label{Sec:spatial_distro}
Studies of the radial distribution of LMXBs in the past yielded apparently
contrasting results. For instance, \cite{Kim03} and \cite{Humph04} found 
that LMXBs follow the host galaxy light distribution in NGC~1316 and 
NGC~1332, while
in NGC~4472 \cite{Kundu02} find that LMXBs follow the GC distribution
better than the optical light. Also, \cite{Sarazin03} and \cite{Jordan04} found no
difference between the distribution of field- and GC-LMXBs, while
\cite{Kundu07} observe that field-LMXBs are more centrally concentrated
than GC-LMXBs. We note, however, that an accurate comparison of these
results is hampered by the different spatial resolutions and
galactocentric ranges covered by these studies. Furthermore, the interplay
between the spatial and color distribution of the GC sample may explain
the discrepancies, since blue and red GCs have different radial
distributions and host different fractions of the LMXB population.

In the case of NGC~1399, we find that the radial distribution of 
field-LMXBs out to $\sim\!5.2\, R_{\rm eff}$ of the diffuse galaxy light, is
significantly steeper than GC-LMXBs, with a statistical significance
$>99.9\%$, in agreement with the results of \cite{Kundu07} for a sample of
five early-type galaxies, however, with smaller galactocentric coverage.
This result supports the conclusion that field LMXBs are not likely to be
formed in GCs and later expelled by three- and four-body interactions, as proposed by e.g. \citet{white02},
since in such a
case field- and GC-LMXBs would be expected to have similar radial surface
density profiles. Field LMXBs (Figure~\ref{rad_prof}, left panel)
follow, in fact, the surface brightness of their host galaxy, suggesting an
evolutionary connection to the main stellar body of the galaxy rather than
to GCs. GC destruction has been proposed as an alternative mechanism to
produce field LMXBs; in this scenario the increased strength of tidal
fields in the galaxy core would result in a field LMXB distribution more
centrally concentrated than the one of their parent GC population if they
were preferentially formed in GCs that moved through the galaxy core 
regions. The effect of the tidal field of the host galaxy on the GC population is possibly
observed in the $\sim\!20\%$ smaller sizes of the total GC populations
(blue and red GCs) within the central 10 kpc ($100\arcsec$, see P11). 
Note, however, that given the low fraction of GC-hosting LMXBs and its dependence
on the host GC color, the production of all field LMXBs in GC requires 
the disruption of a very large fraction of stellar systems, and a very finely tuned
interplay between the original host GC system, color and spatial distribution
to reproduce the observed distributions. 

On the other hand, we do not detect any significant difference between
the GC-LMXB distribution and the overall GC distribution, at odds with
the result of \cite{Kim06} that both red and blue GC-LMXBs
have steeper profiles than blue and red GC populations. Our data suggest
that LMXBs hosted by red GCs simply follow their parent distribution;
the statistics are too low for blue LMXBs to draw significant conclusions,
but we note that an opposite trend is found with LMXBs in blue GCs having 
a shallower number surface density profile around NGC 1399 than the 
corresponding blue GC sub-population.
It is possible that the
\cite{Kim06} result is affected by contamination issues at large radii
due to the lower spatial resolution of their ground-based data, as well
as the fact that NGC~1399 has different properties than the other
galaxies included in their sample.

\begin{figure}[]
\includegraphics[width=8.5cm]{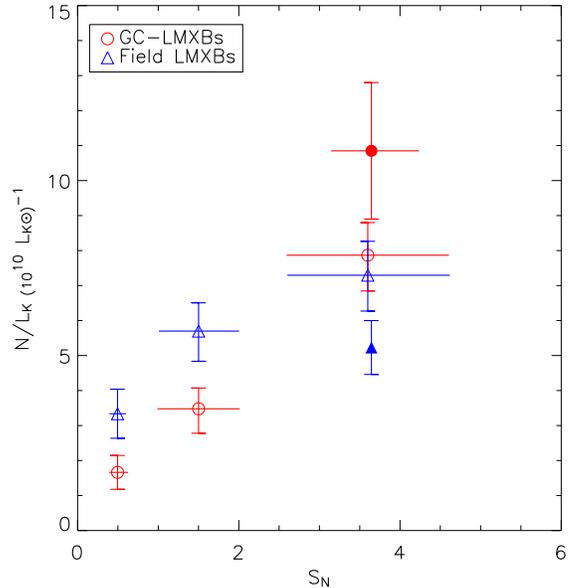}
\caption{Number of LMXBs (normalized to galaxy K band luminosity) as a function of specific frequency $S_N$. Empty symbols represent the \citet{Kim09} dataset, i.e. from left to right NGC 3379, NGC 4697 and NGC 4278, while solid ones represents the values for NGC 1399 derived in this work. All values were calculated within $D_{25}$ and excluding the central 10'' region; the errorbars for NGC1399 include both statistical and systematic uncertainties due to the use of different X-ray LF (see discussion in text). The GC-LMXB abundance is more strongly dependent on $S_N$ than for field source, supporting different evolutionary paths of the two populations.}
\label{Kim09_conf}
\end{figure}

The conclusion that field LMXB follow a different formation path from GC-LMXB
in supported by the results of \citet{Kim09}, based on the analysis of
HST and {\it Chandra} data of 3 nearby ellipticals, showing that the abundance of field LMXBs  
does not depend on the GC specific frequency $S_N$ as strongly as observed for GC-LMXB.
To compare NGC 1399 with the sample of \citet{Kim09} we calculated the number of field and GC LMXBs
within $D_{25}$, as reported by RC3, normalized by total K band luminosity; since \citet{Kim09} reach fainter X-ray luminosities ($L_X>10^{37}$ erg/s) than probed by our data, we extrapolated our LF down to the same limit
using the different LF fits proposed by the latter authors, consisting in either a single or broken power-law,
obtaining correction factors ranging from 1.6 up to 2.3. 
The NGC 1399 GC specific frequency was calculated within the same region (since $S_N$ is known to depend on galactocentric distance, \citealp[see e.g.][]{Dirsch03}), and turns out to be $\sim 30\%$ lower than the global value of 5.1 usually reported in literature.
The result, shown in Figure \ref{Kim09_conf}, supports the difference between GC and field LMXB population, with the former more strongly dependent on $S_N$. The claim that NGC~1399 has an unusually rich LMXB population, down to such faint X-ray luminosities,
depends critically on the assumed LF correction: while using the average correction term would support the peculiar nature of NGC1399, if the faint end of the GC-LMXB LF is flatter than the field one the NGC1399 GC-LMXB population would be comparable (lower errorbar limit in Figure \ref{Kim09_conf}) within the statistical uncertainties to the one of NGC4278.  
We speculate that, if confirmed, the very red GC subpopulation (see $\S$\ref{rad_dep}) could be responsible for part of this difference given the strong effect that metallicity has on LMXB formation; we also point out that although all galaxies used for this plot are giant ellipticals, only NGC 1399 is a central cluster cD galaxy.

On the other hand, our results point toward a number of field binaries consistent with those of poorer
GC systems implying that only a small fraction of LMXB is likely to have escaped from their host GC.

\subsection{The Influence of GC Structure on LMXB Formation}
\label{Disc:struct}
Our analysis of the structural properties of GCs hosting LMXBs (XGC) confirms the
results of previous studies that focused on the core regions of early-type
galaxies \citep{Jordan04, Jordan07a, Sivakoff07}, which found that LMXBs
are preferentially formed in the most compact GCs. The XGC size
distribution is, however, similar to the one of the red GC
sub-population and is $\sim\!20\%$ more compact than the blue GC
sub-population. Thus, we cannot exclude with certainty that the observed
size difference reflects differences in the parent GC population. 
\cite{Spitler06}, in their study of the Sombrero galaxy (NGC 4594), noted that
if there is a radial gradient in GC sizes, projection effects may be responsible
for the observed size difference between red and blue subpopulations
due to their different spatial distribution. Our results however show that
the size difference between red and blue GCs extends out to large galactocentric
radii where projection effects are less pronounced, thus arguing in favour of an intrinsic
effect on the LMXB formation efficiency.

In general, LMXBs prefer high encounter-rate systems,
independent of their host GC  color, as expected if the formation of close
binaries is favoured in high density environments. 

However we observe that GC mass still plays a significant role even
after removing the dependence on central surface brightness and encounter
rate. In particular, while in less massive GCs our findings are in agreement
with those of \citet{Jordan07a, Sivakoff07, Peacock09},
at odds with the latter works is the fact that GCs with luminosities $m_V\!\la\!22$ mag ($M_V\!\la\!
-9.5$ mag) preferentially host LMXBs independent of their structural
properties. We point out that this result is confirmed also if we limit the analysis to the central
$2'$ from the galaxy center and is thus not affected by contamination problems which would be important mainly in the galaxy outskirts.
This result could be explained if massive clusters are
more efficient in retaining a larger fraction of their neutron stars
\citep{Verbunt05}. However \cite{Smits06}, evaluating a range of possible SNe kick distributions and GC potentials, conclude that in most cases the estimated retention fraction produces a simple linear dependence of the likelihood to find a LMXB on GC mass. A possible solution would be to assume that if SN kicks follow a bimodal distribution peaked around 10 and 200 km/s, as proposed as one of the most likely scenarios by the latter authors, the most massive GCs are able to retain some of the fast neutron stars, thus disrupting the otherwise linear dependence on GC mass.
In such case it is possible that previous studies have missed this effect due to, e.g., the binning adopted in \cite{Smits06}, or the small number of extremely bright clusters found in M31 by \cite{Peacock09}.

While it is indeed difficulty to measure structural parameters for GCs more distant than 10 Mpc, we point out that our results are based on the relative differences between subsamples within the same dataset, which are more robust than the absolute values. Furthermore, following \cite{Sivakoff07}, we did test our results using the more robust effective radius, instead of the core radius, to compute the interaction rate $\Gamma$.
We find no significant change in the results, with the exception of Figure 16, where the difference in interaction rate between the global GC population and X-ray GCs as a function of magnitude becomes smaller. This would enhance the contrast with previous results, in making structural parameters even less important for LMXB formation. We conclude that the 
collision rates computed using the estimated core radii have more 
predictive power for whether a cluster will contain an X-ray source than 
collision rates computed using the half-light radius.  This is 
unsurprising, since the core radii and half-light radii are not well 
correlated in globular clusters, and the bulk of dynamical interactions 
take place within the cluster cores.

\subsection{LMXB properties}
The X-ray properties (luminosity function, hardness ratios) of the LMXB
population are in fair agreement with the literature, but we observe a
marginally steeper LF for field LMXBs \citep[see also][]{Kundu07}. We do
not observe any correlation between LMXB properties (luminosity,
spectral hardness, variability) and those of the host GCs, indicating
that the LMXB evolution (if not the formation) is primarily driven by
the properties of the stellar binary system and not of the host GC. Our temporal analysis supports the view that most of the X-ray emission from GCs 
is produced by a single accreting binary since a significant fraction of the bright LMXBs shows signs of variability, as do LMXBs in the more massive GCs. On the other hand the steeper LF of field LMXBs and the brighter median LMXB luminosity observed in the galaxy center, suggest that some of the brighest GCs may harbor multiple accreting binaries (if we assume that all LMXBs share the same intrinsic LF), in agreement with the conclusions of, e.g., \cite{Kundu07} based on a larger sample of early-type galaxies or \cite{Fabbiano10} for NGC 4278, and as observed in the Galactic GC M15 by \cite{White01}.

In particular the
brightest color-confirmed GC X-ray source ($L_X\sim 4\times 10^{39}$ erg s$^{-1}$), which
resides in one of the most metal-rich GC, does not exhibit signs of
variability, possibly indicating the presence of multiple accreting
X-ray binaries. Thus we cannot confirm the presence of intermediate-mass black holes with average $L_X>10^{39}~\mbox{erg s}^{-1}$, as reported by \cite{Irwin10}. This is not in contradiction with the latter study however, since our color catalog does not include their source and thus we could not confirm its GC nature based only on the data utilized here.

\section{Conclusions}
We perform the first wide-field, high spatial resolution study of the
LMXB/GC connection in the massive early-type galaxy NGC~1399 covering
galactocentric distances out to $\sim 50$ kpc ($> 5~R_{\rm eff}$). Our 
analysis reveals the following key results:\\

\noindent$\bullet$ NGC 1399 has the highest fraction of LMXBs residing in GCs of all early-type galaxies studied so far, $f_{\rm GC-LMXB}=65\%\pm5\%$, even after accounting for its rich GC system and large stellar mass.\\

\noindent$\bullet$ The LMXB fraction depends on galactocentric distance since the distributions of field- and GC-LMXBs follow different radial surface density profiles. This argues against a common origin of all LMXBs.\\

\noindent$\bullet$ The majority of LMXBs are hosted by the red GC population, which closely follows the optical galaxy light, while the blue GC-LMXB population has a more extended profile. We also confirm the presence of a very red GC sub-population residing in the galaxy core that hosts a large fraction of LMXBs.\\

\noindent$\bullet$ We find that LMXBs tend to follow the spatial distribution of the red GC sub-population, thus suggesting that dynamical interactions of GCs with the host galaxy do not affect the LMXB formation.\\

\noindent$\bullet$ GC mass, color (metallicity), and interaction rate $\Gamma$ all seem to affect the LMXB formation likelihood at any given galactocentric distance.\\

\noindent$\bullet$ We find no evidence of a dependence of LMXB properties on those of the host GCs, as expected if LMXB evolution is primarily driven by the properties of stellar binary systems. While most GCs are likely to host a single LMXB, the steeper LF of field-LMXB, the higher median X-ray luminosity of GC-LMXBs and the lack of variability in the brightest color-confirmed GC X-ray source, support the presence of multiple accreting binaries in some of the X-ray brightest GCs.\\


Hopefully the restored HST/ACS capabilities will allow more wide-field
studies of nearby early-type galaxies and their GC systems, in order to 
extend the present results and finally remove the degeneracies between 
galactocentric distance and GC color as well as GC structural parameters 
which, so far, have hampered a proper understanding of the physical 
processes driving LMXB formation and evolution.

\acknowledgments




{\it Facilities:} \facility{HST (ACS)}, \facility{CXO (ACIS)}.



\begin{acknowledgments}
We thank E. Flaccomio, A. Zezas for useful suggestions and comments,
and T. Richtler for providing access to his ground-based photometric
catalogs. T.H.P. acknowledges financial support in form of the Plaskett
Research Fellowship at the Herzberg Institute of Astrophysics of the
National Research Council of Canada. He is also grateful for the support
and warm hospitality at the Federico II University of Naples where parts
of this work were conducted. We thank the anonymous referee for many
useful suggestions.\\
MP acknowledges support from the ASI-INAF contract I/009/10/0.
AK acknowledges support from HST archival program HST-AR-11264.
SEZ acknowledges support for this work from the NASA ADP grant
NNX08AJ60G\\
Support for {\it HST\/} Program GO-10129  was provided by NASA through a grant from the Space
Telescope Science Institute, which is operated by the Association of
Universities for Research in Astronomy, Inc., under NASA contract
NAS5--26555.\end{acknowledgments}

\clearpage
\LongTables

\begin{landscape}
\begin{deluxetable}{ccccccccccccccc}
\tablecolumns{15} 
\tabletypesize{\scriptsize}
\tablecaption{X-ray source catalog \label{tbl-1}}
\tablewidth{0pt}
\tablehead{
\colhead{ID} & \colhead{RA} & \colhead{DEC} & \colhead{Net counts} & \colhead{Flux ($10^{-7}$ ph/s/cm$^2$)} & \colhead{HR1} & \colhead{HR2} & \colhead{Var.} & \colhead{X-opt sep.} & \colhead{$V$} & Stellar & \colhead{$T1$} & \colhead{$C-T1$} & $z$ & $g-z$\\
 & \multicolumn{2}{c}{(J2000)} & \colhead{(0.5-8 keV)} &\colhead{(0.5-8 keV)} & \colhead{(0.5-1 vs 1-2 keV)} & \colhead{(0.5-1 vs 2-8 keV)} & & \colhead{(arcsec)} & \colhead{(F606V)} & index &  & & & \\}
\startdata
  1 &  3:38:10.4 & -35:27:59.8 & $  75.3_{-10.5}^{+ 9.5}$ & $  25.8\pm  3.4$ & $ -0.02\pm   0.20$ & $  0.49\pm   0.19$ & 0 & $0.16$ & $23.8$ & $1.0$ & \nodata & \nodata & \nodata & \nodata \\
  2 &  3:38:11.6 & -35:26:49.0 & $  97.2_{-11.7}^{+10.6}$ & $  23.6\pm  2.7$ & $  0.01\pm   0.14$ & $ -0.18\pm   0.17$ & 0 & $1.73$ & $24.7$ & $1.0$ & \nodata & \nodata & \nodata & \nodata \\
  3 &  3:38:12.0 & -35:27: 0.4 & $  54.6_{- 9.5}^{+ 8.4}$ & $  15.5\pm  2.5$ & $  0.50\pm   0.33$ & $  0.75\pm   0.27$ & 0 & $0.01$ & $21.0$ & $0.9$ & \nodata & \nodata & \nodata & \nodata \\
  4 &  3:38:12.7 & -35:28:57.3 & $  85.8_{-10.9}^{+ 9.8}$ & $  26.1\pm  3.1$ & $  0.05\pm   0.16$ & $ -0.08\pm   0.18$ & 0 & $0.27$ & $20.8$ & $0.9$ & $20.2$ & $ 1.61$ & \nodata & \nodata \\
  5 &  3:38:12.8 & -35:25:19.5 & $  33.1_{- 8.8}^{+ 7.8}$ & $  12.4\pm  3.1$ & $ -0.19\pm   0.33$ & $  0.13\pm   0.36$ & 0 & $4.37$ & $27.7$ & $0.5$ & \nodata & \nodata & \nodata & \nodata \\
  6 &  3:38:12.9 & -35:25:59.8 & $  30.7_{- 8.5}^{+ 7.4}$ & $   6.6\pm  1.7$ & $  0.19\pm   0.38$ & $ -0.01\pm   0.43$ & 0 & $0.23$ & $22.2$ & $0.9$ & \nodata & \nodata & \nodata & \nodata \\
  7 &  3:38:14.7 & -35:24:28.8 & $  20.6_{- 7.3}^{+ 6.3}$ & $   8.6\pm  2.8$ & $  0.12\pm   0.41$ & $  0.35\pm   0.61$ & 0 & $0.66$ & $27.8$ & $0.6$ & \nodata & \nodata & \nodata & \nodata \\
  8 &  3:38:15.0 & -35:23:58.3 & $  23.8_{- 7.7}^{+ 6.6}$ & $  10.5\pm  3.1$ & $  0.11\pm   0.38$ & $  0.61\pm   0.57$ & 0 & $1.41$ & $27.9$ & $0.1$ & \nodata & \nodata & \nodata & \nodata \\
  9 &  3:38:15.1 & -35:27:57.7 & $  38.7_{- 8.1}^{+ 7.0}$ & $  10.6\pm  2.1$ & $ -0.02\pm   0.27$ & $  0.06\pm   0.29$ & 0 & $0.23$ & $22.0$ & $0.9$ & $21.2$ & $ 1.74$ & \nodata & \nodata \\
 10 &  3:38:15.5 & -35:26:29.4 & $  86.9_{-11.2}^{+10.1}$ & $  21.9\pm  2.7$ & $ -0.31\pm   0.14$ & $ -1.02\pm   0.25$ & 1 & $3.92$ & $25.1$ & $0.2$ & \nodata & \nodata & \nodata & \nodata \\
 11 &  3:38:15.6 & -35:26: 0.8 & $  12.8_{- 6.4}^{+ 5.3}$ & $   5.6\pm  2.6$ & $ -0.12\pm   0.70$ & $  0.57\pm   0.63$ & 0 & $0.68$ & $22.0$ & $1.0$ & $21.3$ & $ 1.79$ & \nodata & \nodata \\
 12 &  3:38:16.3 & -35:26:33.2 & $  24.7_{- 7.3}^{+ 6.2}$ & $   6.7\pm  1.8$ & $ -0.03\pm   0.37$ & $  0.23\pm   0.36$ & 0 & $2.07$ & $27.3$ & $1.0$ & \nodata & \nodata & \nodata & \nodata \\
 13 &  3:38:16.4 & -35:27:33.8 & $  21.5_{- 6.9}^{+ 5.8}$ & $   9.1\pm  2.7$ & $ -0.62\pm   0.46$ & $  0.13\pm   0.35$ & 0 & $0.20$ & $22.8$ & $1.0$ & $22.0$ & $ 1.66$ & \nodata & \nodata \\
 14 &  3:38:16.5 & -35:27:45.7 & $  56.0_{- 9.1}^{+ 8.1}$ & $  15.9\pm  2.4$ & $  0.01\pm   0.20$ & $  0.26\pm   0.22$ & 0 & $0.18$ & $22.6$ & $1.0$ & $22.1$ & $ 1.01$ & \nodata & \nodata \\
 15 &  3:38:16.8 & -35:26:14.5 & $  20.0_{- 6.9}^{+ 5.8}$ & $   8.0\pm  2.5$ & $  0.25\pm   0.72$ & $  0.81\pm   0.46$ & 0 & $0.21$ & $23.3$ & $1.0$ & $22.5$ & $ 1.87$ & \nodata & \nodata \\
 16 &  3:38:17.2 & -35:27:33.7 & $  20.3_{- 6.7}^{+ 5.6}$ & $   4.6\pm  1.4$ & $ -0.29\pm   0.39$ & $ -0.31\pm   0.40$ & 0 & $0.72$ & $23.0$ & $1.0$ & $22.1$ & $ 1.78$ & \nodata & \nodata \\
 17 &  3:38:17.4 & -35:28: 7.7 & $  29.0_{- 7.3}^{+ 6.3}$ & $   7.5\pm  1.8$ & $ -0.03\pm   0.34$ & $  0.16\pm   0.32$ & 0 & $1.08$ & $26.6$ & $0.8$ & \nodata & \nodata & \nodata & \nodata \\
 18 &  3:38:17.6 & -35:23:51.3 & $  23.5_{- 8.0}^{+ 6.9}$ & $   8.2\pm  2.6$ & $ -0.19\pm   0.47$ & $  0.24\pm   0.42$ & 0 & $4.92$ & $24.1$ & $1.0$ & \nodata & \nodata & \nodata & \nodata \\
 19 &  3:38:17.8 & -35:27:50.6 & $  16.8_{- 6.1}^{+ 5.0}$ & $   5.4\pm  1.8$ & $  0.01\pm   0.37$ & $  0.17\pm   0.77$ & 0 & $0.12$ & $22.7$ & $1.0$ & $21.9$ & $ 1.56$ & \nodata & \nodata \\
 20 &  3:38:18.9 & -35:27:32.5 & $  73.7_{-10.1}^{+ 9.1}$ & $  19.9\pm  2.6$ & $ -0.19\pm   0.17$ & $ -0.03\pm   0.17$ & 0 & $0.32$ & $23.4$ & $1.0$ & \nodata & \nodata & \nodata & \nodata \\
 21 &  3:38:18.9 & -35:28: 2.5 & $  37.6_{- 7.9}^{+ 6.8}$ & $   9.2\pm  1.8$ & $ -0.29\pm   0.22$ & $ -1.09\pm   0.43$ & 0 & $4.72$ & $27.7$ & $0.0$ & \nodata & \nodata & \nodata & \nodata \\
 22 &  3:38:19.3 & -35:26:11.0 & $  30.7_{- 7.8}^{+ 6.7}$ & $   7.1\pm  1.7$ & $ -0.36\pm   0.28$ & $ -1.06\pm   0.51$ & 0 & $0.24$ & $22.3$ & $1.0$ & $21.5$ & $ 1.74$ & \nodata & \nodata \\
 23 &  3:38:19.3 & -35:27:34.4 & $  30.7_{- 7.3}^{+ 6.2}$ & $   7.1\pm  1.6$ & $ -0.08\pm   0.28$ & $ -0.36\pm   0.35$ & 0 & $0.85$ & $23.5$ & $1.0$ & \nodata & \nodata & \nodata & \nodata \\
 24 &  3:38:19.5 & -35:25: 0.2 & $  21.5_{- 7.4}^{+ 6.3}$ & $   7.1\pm  2.3$ & $  0.08\pm   0.49$ & $  0.40\pm   0.44$ & 0 & $0.65$ & $24.1$ & $1.0$ & $23.2$ & $ 1.77$ & \nodata & \nodata \\
 25 &  3:38:19.7 & -35:29:36.8 & $  12.7_{- 5.5}^{+ 4.3}$ & $   3.8\pm  1.5$ & $ -0.07\pm   0.51$ & $  0.03\pm   0.50$ & 0 & $0.60$ & $24.9$ & $1.0$ & $24.0$ & $ 0.81$ & \nodata & \nodata \\
 26 &  3:38:19.8 & -35:28:46.0 & $  62.3_{- 9.4}^{+ 8.3}$ & $  20.5\pm  2.9$ & $ -0.41\pm   0.17$ & $ -0.66\pm   0.22$ & 0 & $2.90$ & $21.0$ & $0.0$ & \nodata & \nodata & \nodata & \nodata \\
 27 &  3:38:20.0 & -35:26:43.8 & $  22.8_{- 7.1}^{+ 6.0}$ & $   5.2\pm  1.5$ & $ -0.02\pm   0.44$ & $  0.04\pm   0.41$ & 0 & $0.30$ & $21.4$ & $1.0$ & $20.6$ & $ 1.51$ & \nodata & \nodata \\
 28 &  3:38:20.1 & -35:24:46.9 & $ 194.5_{-15.6}^{+14.6}$ & $  74.5\pm  5.8$ & $ -0.19\pm   0.09$ & $ -0.14\pm   0.11$ & 1 & $0.47$ & $21.6$ & $0.0$ & \nodata & \nodata & \nodata & \nodata \\
 29 &  3:38:20.2 & -35:28:31.1 & $  24.2_{- 6.7}^{+ 5.6}$ & $   9.5\pm  2.4$ & $  0.64\pm   0.62$ & $  0.89\pm   0.41$ & 0 & $0.22$ & $22.3$ & $0.0$ & \nodata & \nodata & \nodata & \nodata \\
 30 &  3:38:20.4 & -35:29:28.2 & $  15.1_{- 5.7}^{+ 4.6}$ & $   4.7\pm  1.6$ & $ -0.11\pm   0.39$ & $ -0.60\pm   0.63$ & 0 & $0.16$ & $21.5$ & $0.9$ & $20.7$ & $ 1.56$ & \nodata & \nodata \\
 31 &  3:38:20.8 & -35:27:26.5 & $  32.5_{- 7.5}^{+ 6.4}$ & $  11.1\pm  2.4$ & $ -0.64\pm   0.30$ & $ -0.37\pm   0.26$ & 0 & $0.46$ & $21.1$ & $0.9$ & \nodata & \nodata & \nodata & \nodata \\
 32 &  3:38:20.9 & -35:24:57.1 & $  30.0_{- 7.8}^{+ 6.7}$ & $   9.2\pm  2.2$ & $  0.13\pm   0.32$ & $  0.13\pm   0.36$ & 0 & $0.46$ & $23.1$ & $0.9$ & $22.3$ & $ 1.99$ & \nodata & \nodata \\
 33 &  3:38:21.0 & -35:27: 2.6 & $  40.9_{- 8.2}^{+ 7.1}$ & $  12.0\pm  2.3$ & $ -0.02\pm   0.23$ & $  0.04\pm   0.27$ & 0 & $0.34$ & $23.1$ & $1.0$ & $22.2$ & $ 2.20$ & \nodata & \nodata \\
 34 &  3:38:21.0 & -35:27:24.5 & $  44.1_{- 8.4}^{+ 7.3}$ & $  12.1\pm  2.1$ & $ -0.14\pm   0.20$ & $ -0.52\pm   0.28$ & 0 & $0.47$ & $22.7$ & $1.0$ & $21.9$ & $ 2.15$ & \nodata & \nodata \\
 35 &  3:38:21.0 & -35:30:13.2 & $  39.2_{- 8.0}^{+ 6.9}$ & $  10.5\pm  2.0$ & $ -0.14\pm   0.26$ & $ -0.10\pm   0.26$ & 0 & $0.47$ & $21.6$ & $0.9$ & $20.9$ & $ 1.39$ & \nodata & \nodata \\
 36 &  3:38:21.1 & -35:27:32.6 & $  37.4_{- 7.8}^{+ 6.7}$ & $  10.8\pm  2.1$ & $  0.05\pm   0.25$ & $ -0.03\pm   0.28$ & 0 & $0.34$ & $24.1$ & $1.0$ & \nodata & \nodata & \nodata & \nodata \\
 37 &  3:38:21.6 & -35:27:43.4 & $  21.4_{- 6.5}^{+ 5.4}$ & $   6.9\pm  1.9$ & $  0.33\pm   0.58$ & $  0.43\pm   0.54$ & 0 & $2.36$ & $23.5$ & $0.9$ & \nodata & \nodata & \nodata & \nodata \\
 38 &  3:38:21.7 & -35:26:36.4 & $  20.5_{- 6.6}^{+ 5.5}$ & $   7.2\pm  2.1$ & $ -0.10\pm   0.47$ & $  0.40\pm   0.39$ & 0 & $0.71$ & $22.3$ & $1.0$ & \nodata & \nodata & \nodata & \nodata \\
 39 &  3:38:21.9 & -35:24:21.9 & $  14.5_{- 7.0}^{+ 6.0}$ & $   2.1\pm  0.9$ & $  0.17\pm   0.53$ & $ -4.68\pm  \infty$ & 1 & $0.11$ & $25.4$ & $0.8$ & \nodata & \nodata & \nodata & \nodata \\
 40 &  3:38:21.9 & -35:29:29.1 & $ 227.6_{-16.4}^{+15.3}$ & $  64.7\pm  4.5$ & $ -0.01\pm   0.09$ & $  0.12\pm   0.10$ & 0 & $0.24$ & $24.6$ & $1.0$ & \nodata & \nodata & \nodata & \nodata \\
 41 &  3:38:22.1 & -35:29: 1.1 & $  15.3_{- 5.9}^{+ 4.8}$ & $   5.2\pm  1.8$ & $  0.74\pm   0.63$ & $  0.84\pm   0.76$ & 0 & $3.37$ & $27.4$ & $0.0$ & \nodata & \nodata & \nodata & \nodata \\
 42 &  3:38:22.2 & -35:28:41.3 & $  15.5_{- 5.9}^{+ 4.8}$ & $   6.3\pm  2.2$ & $ -0.37\pm   0.42$ & $ -0.04\pm   0.57$ & 0 & $4.50$ & $25.9$ & $1.0$ & \nodata & \nodata & \nodata & \nodata \\
 43 &  3:38:22.5 & -35:29:52.1 & $  25.6_{- 6.8}^{+ 5.7}$ & $   6.7\pm  1.6$ & $ -0.41\pm   0.35$ & $ -0.32\pm   0.32$ & 0 & $0.22$ & $21.9$ & $1.0$ & \nodata & \nodata & \nodata & \nodata \\
 44 &  3:38:22.6 & -35:27:52.7 & $  14.2_{- 5.9}^{+ 4.8}$ & $   3.7\pm  1.4$ & $ -0.50\pm   0.48$ & $ -1.01\pm   0.73$ & 0 & $1.94$ & $25.3$ & $1.0$ & \nodata & \nodata & \nodata & \nodata \\
 45 &  3:38:22.7 & -35:28:48.4 & $  14.8_{- 5.9}^{+ 4.8}$ & $   4.3\pm  1.5$ & $ -0.41\pm   0.50$ & $ -0.27\pm   0.46$ & 0 & $0.34$ & $22.5$ & $1.0$ & $21.7$ & $ 1.92$ & \nodata & \nodata \\
 46 &  3:38:23.0 & -35:25: 3.9 & $  24.5_{- 7.0}^{+ 5.9}$ & $  12.4\pm  3.3$ & $ -0.44\pm   0.33$ & $ -0.20\pm   0.40$ & 0 & $1.15$ & $22.5$ & $1.0$ & \nodata & \nodata & \nodata & \nodata \\
 47 &  3:38:23.2 & -35:27:10.0 & $  54.9_{- 9.2}^{+ 8.1}$ & $  15.6\pm  2.5$ & $ -0.24\pm   0.20$ & $ -0.07\pm   0.21$ & 0 & $1.32$ & $24.7$ & $1.0$ & \nodata & \nodata & \nodata & \nodata \\
 48 &  3:38:23.2 & -35:27:14.8 & $  22.4_{- 6.9}^{+ 5.8}$ & $   5.5\pm  1.6$ & $  0.21\pm   0.46$ & $  0.02\pm   0.56$ & 0 & $0.44$ & $22.4$ & $1.0$ & $21.6$ & $ 1.92$ & $20.6$ & $ 2.14$ \\
 49 &  3:38:23.2 & -35:28: 4.0 & $  31.3_{- 7.4}^{+ 6.3}$ & $   7.8\pm  1.7$ & $  0.07\pm   0.28$ & $ -0.68\pm   0.56$ & 0 & $0.12$ & $22.3$ & $1.0$ & \nodata & \nodata & $20.6$ & $ 2.19$ \\
 50 &  3:38:23.4 & -35:28:25.8 & $  34.6_{- 7.6}^{+ 6.6}$ & $   8.1\pm  1.7$ & $  0.17\pm   0.26$ & $ -0.42\pm   0.44$ & 0 & $2.76$ & $27.5$ & $0.4$ & \nodata & \nodata & \nodata & \nodata \\
 51 &  3:38:23.5 & -35:25:57.5 & $  24.6_{- 6.8}^{+ 5.7}$ & $   5.8\pm  1.5$ & $  0.17\pm   0.34$ & $ -0.04\pm   0.40$ & 1 & $0.61$ & $23.3$ & $1.0$ & $22.3$ & $ 2.13$ & \nodata & \nodata \\
 52 &  3:38:24.1 & -35:28:39.7 & $  27.1_{- 7.1}^{+ 6.0}$ & $   9.0\pm  2.2$ & $  0.20\pm   0.68$ & $  0.85\pm   0.34$ & 0 & $1.90$ & $23.0$ & $0.9$ & \nodata & \nodata & \nodata & \nodata \\
 53 &  3:38:24.9 & -35:28: 3.3 & $  28.2_{- 7.2}^{+ 6.1}$ & $   6.6\pm  1.5$ & $ -0.00\pm   0.30$ & $ -0.17\pm   0.35$ & 0 & $0.21$ & $24.6$ & $0.6$ & \nodata & \nodata & $22.9$ & $ 1.89$ \\
 54 &  3:38:25.1 & -35:23:53.0 & $  17.9_{- 7.3}^{+ 6.2}$ & $   3.6\pm  1.4$ & $  1.08\pm   0.82$ & $  1.14\pm   1.05$ & 0 & $0.37$ & $27.3$ & $0.8$ & \nodata & \nodata & \nodata & \nodata \\
 55 &  3:38:25.3 & -35:25:22.7 & $ 256.7_{-17.7}^{+16.6}$ & $  72.7\pm  4.9$ & $ -0.31\pm   0.08$ & $ -0.31\pm   0.09$ & 1 & $0.71$ & $20.6$ & $1.0$ & \nodata & \nodata & $19.3$ & $ 0.77$ \\
 56 &  3:38:25.3 & -35:27:53.8 & $  81.2_{-10.6}^{+ 9.5}$ & $  25.0\pm  3.1$ & $ -0.19\pm   0.16$ & $  0.03\pm   0.17$ & 0 & $0.34$ & $21.9$ & $1.0$ & \nodata & \nodata & $20.1$ & $ 2.11$ \\
 57 &  3:38:25.5 & -35:22:44.7 & $ 160.2_{-14.7}^{+13.7}$ & $  58.9\pm  5.2$ & $ -0.39\pm   0.11$ & $ -0.44\pm   0.12$ & 0 & $0.15$ & $22.8$ & $1.0$ & $21.7$ & $ 0.81$ & \nodata & \nodata \\
 58 &  3:38:25.5 & -35:26:47.3 & $  19.1_{- 7.4}^{+ 6.3}$ & $   6.1\pm  2.2$ & $ -0.85\pm   0.66$ & $ -0.10\pm   0.35$ & 1 & $0.68$ & $25.0$ & $1.0$ & \nodata & \nodata & $23.0$ & $ 1.94$ \\
 59 &  3:38:25.6 & -35:25:56.4 & $  75.9_{-10.6}^{+ 9.5}$ & $  21.1\pm  2.8$ & $  0.09\pm   0.17$ & $  0.30\pm   0.20$ & 0 & $0.88$ & $23.7$ & $1.0$ & $23.2$ & $ 1.65$ & $22.1$ & $ 1.75$ \\
 60 &  3:38:25.6 & -35:26:44.1 & $  25.9_{- 7.9}^{+ 6.8}$ & $   9.1\pm  2.6$ & $ -0.46\pm   0.39$ & $  0.03\pm   0.34$ & 0 & $1.42$ & $25.1$ & $1.0$ & \nodata & \nodata & \nodata & \nodata \\
 61 &  3:38:25.7 & -35:27:30.2 & $  48.6_{- 8.8}^{+ 7.7}$ & $  15.2\pm  2.6$ & $ -0.30\pm   0.22$ & $ -0.16\pm   0.22$ & 1 & $0.31$ & $23.2$ & $1.0$ & \nodata & \nodata & $21.5$ & $ 1.85$ \\
 62 &  3:38:25.8 & -35:24:43.1 & $  15.6_{- 6.3}^{+ 5.2}$ & $   4.8\pm  1.8$ & $ -0.18\pm   0.42$ & $ -0.23\pm   0.69$ & 0 & $1.41$ & $27.5$ & $0.7$ & \nodata & \nodata & \nodata & \nodata \\
 63 &  3:38:25.8 & -35:30:12.9 & $  14.5_{- 5.8}^{+ 4.7}$ & $   3.4\pm  1.2$ & $  0.08\pm   0.48$ & $ -0.54\pm   0.80$ & 0 & $0.20$ & $21.6$ & $0.9$ & $22.2$ & $ 2.04$ & \nodata & \nodata \\
 64 &  3:38:25.9 & -35:27:57.1 & $  70.4_{- 9.9}^{+ 8.9}$ & $  19.6\pm  2.6$ & $  0.09\pm   0.18$ & $  0.22\pm   0.20$ & 0 & $0.31$ & $22.2$ & $1.0$ & $21.5$ & $ 1.96$ & $20.6$ & $ 2.09$ \\
 65 &  3:38:26.0 & -35:27:42.6 & $ 117.6_{-12.3}^{+11.3}$ & $  31.4\pm  3.2$ & $ -0.14\pm   0.12$ & $ -0.22\pm   0.14$ & 0 & $1.70$ & $27.5$ & $0.2$ & \nodata & \nodata & \nodata & \nodata \\
 66 &  3:38:26.1 & -35:24: 0.1 & $  23.4_{- 7.5}^{+ 6.4}$ & $   5.2\pm  1.6$ & $ -0.42\pm   0.32$ & $ -1.39\pm   1.11$ & 0 & $1.76$ & $26.6$ & $0.7$ & \nodata & \nodata & \nodata & \nodata \\
 67 &  3:38:26.2 & -35:27:37.0 & $  17.7_{- 6.5}^{+ 5.4}$ & $   4.0\pm  1.4$ & $  0.73\pm   0.63$ & $  0.82\pm   0.57$ & 0 & $0.40$ & $23.5$ & $1.0$ & \nodata & \nodata & $21.8$ & $ 2.14$ \\
 68 &  3:38:26.4 & -35:26:35.0 & $  27.5_{- 7.8}^{+ 6.7}$ & $   7.5\pm  2.0$ & $ -0.10\pm   0.38$ & $  0.23\pm   0.34$ & 0 & $0.58$ & $22.6$ & $1.0$ & $21.6$ & $ 1.74$ & $20.8$ & $ 2.15$ \\
 69 &  3:38:26.5 & -35:27:32.7 & $ 142.4_{-13.5}^{+12.5}$ & $  37.3\pm  3.4$ & $ -0.09\pm   0.11$ & $ -0.18\pm   0.13$ & 1 & $1.91$ & $26.6$ & $1.0$ & \nodata & \nodata & \nodata & \nodata \\
 70 &  3:38:26.6 & -35:27:12.2 & $  31.7_{- 8.4}^{+ 7.3}$ & $  12.2\pm  3.0$ & $ -0.43\pm   0.30$ & $ -0.49\pm   0.37$ & 1 & $0.17$ & $22.3$ & $1.0$ & \nodata & \nodata & $20.4$ & $ 2.21$ \\
 71 &  3:38:26.7 & -35:27: 5.2 & $  32.7_{- 8.9}^{+ 7.9}$ & $   7.5\pm  1.9$ & $  0.26\pm   0.31$ & $ -0.57\pm   0.65$ & 0 & $1.00$ & $23.1$ & $1.0$ & \nodata & \nodata & $21.6$ & $ 1.83$ \\
 72 &  3:38:26.7 & -35:27: 9.5 & $  38.7_{- 8.9}^{+ 7.9}$ & $  11.5\pm  2.5$ & $ -0.36\pm   0.25$ & $ -0.88\pm   0.39$ & 0 & $0.32$ & $22.8$ & $1.0$ & \nodata & \nodata & $21.0$ & $ 2.04$ \\
 73 &  3:38:27.0 & -35:27: 4.6 & $  17.9_{- 9.5}^{+ 8.4}$ & $   4.8\pm  2.4$ & $ -0.13\pm   0.61$ & $ -0.94\pm   1.31$ & 0 & $1.74$ & $25.3$ & $1.0$ & \nodata & \nodata & \nodata & \nodata \\
 74 &  3:38:27.0 & -35:29:46.0 & $  28.3_{- 7.0}^{+ 5.9}$ & $   6.7\pm  1.5$ & $  0.10\pm   0.26$ & $ -0.90\pm   0.61$ & 0 & $0.19$ & $24.5$ & $1.0$ & \nodata & \nodata & \nodata & \nodata \\
 75 &  3:38:27.1 & -35:25:57.6 & $  23.2_{- 7.4}^{+ 6.3}$ & $   6.4\pm  1.9$ & $ -0.40\pm   0.42$ & $ -0.48\pm   0.45$ & 0 & $1.20$ & $24.9$ & $0.9$ & \nodata & \nodata & $23.0$ & $ 2.21$ \\
 76 &  3:38:27.2 & -35:26: 1.8 & $  98.8_{-11.9}^{+10.8}$ & $  25.5\pm  2.9$ & $  0.05\pm   0.15$ & $  0.12\pm   0.17$ & 1 & $0.87$ & $21.8$ & $0.9$ & $21.0$ & $ 2.11$ & $20.1$ & $ 2.17$ \\
 77 &  3:38:27.3 & -35:27:17.1 & $  41.7_{- 9.6}^{+ 8.5}$ & $  11.8\pm  2.6$ & $  0.44\pm   0.39$ & $  0.43\pm   0.43$ & 0 & $1.85$ & $26.6$ & $0.9$ & \nodata & \nodata & \nodata & \nodata \\
 78 &  3:38:27.5 & -35:30:34.3 & $  26.9_{- 7.1}^{+ 6.0}$ & $   6.2\pm  1.5$ & $ -0.15\pm   0.35$ & $ -0.10\pm   0.33$ & 0 & $0.32$ & $24.1$ & $1.0$ & $23.2$ & $ 1.65$ & \nodata & \nodata \\
 79 &  3:38:27.6 & -35:26: 5.8 & $  70.5_{-10.4}^{+ 9.3}$ & $  19.7\pm  2.7$ & $ -0.07\pm   0.18$ & $  0.03\pm   0.20$ & 1 & $0.63$ & $22.2$ & $0.9$ & $21.3$ & $ 1.99$ & $20.6$ & $ 2.13$ \\
 80 &  3:38:27.7 & -35:26:49.0 & $ 509.8_{-25.2}^{+24.1}$ & $ 135.1\pm  6.5$ & $ -0.33\pm   0.06$ & $ -0.70\pm   0.08$ & 0 & $1.40$ & $23.4$ & $1.0$ & \nodata & \nodata & \nodata & \nodata \\
 81 &  3:38:27.8 & -35:25:27.1 & $  75.5_{-10.5}^{+ 9.4}$ & $  21.5\pm  2.8$ & $  0.00\pm   0.20$ & $  0.31\pm   0.19$ & 0 & $2.66$ & $22.5$ & $1.0$ & \nodata & \nodata & \nodata & \nodata \\
 82 &  3:38:27.8 & -35:27:51.1 & $ 107.0_{-12.2}^{+11.1}$ & $  29.5\pm  3.2$ & $  0.26\pm   0.17$ & $  0.51\pm   0.16$ & 0 & $0.37$ & $24.6$ & $0.9$ & \nodata & \nodata & $23.0$ & $ 1.82$ \\
 83 &  3:38:27.9 & -35:27:42.4 & $  24.0_{- 7.5}^{+ 6.4}$ & $   5.7\pm  1.6$ & $ -0.47\pm   0.34$ & $ -0.92\pm   0.48$ & 0 & $3.18$ & $26.3$ & $0.1$ & \nodata & \nodata & \nodata & \nodata \\
 84 &  3:38:27.9 & -35:27:47.6 & $  72.3_{-10.4}^{+ 9.4}$ & $  19.0\pm  2.6$ & $  0.17\pm   0.18$ & $  0.19\pm   0.21$ & 0 & $2.52$ & $24.6$ & $1.0$ & \nodata & \nodata & $24.1$ & $ 2.85$ \\
 85 &  3:38:28.1 & -35:25:45.9 & $  26.3_{- 7.5}^{+ 6.5}$ & $   7.0\pm  1.8$ & $ -0.60\pm   0.44$ & $ -0.14\pm   0.29$ & 0 & $0.99$ & $22.5$ & $0.8$ & $21.6$ & $ 1.88$ & $20.8$ & $ 2.11$ \\
 86 &  3:38:28.2 & -35:22: 5.0 & $  15.1_{- 6.4}^{+ 5.3}$ & $   4.9\pm  1.9$ & $  0.78\pm   0.67$ & $  0.51\pm   1.02$ & 0 & $6.92$ & $26.4$ & $0.3$ & \nodata & \nodata & \nodata & \nodata \\
 87 &  3:38:28.2 & -35:25:51.2 & $  48.8_{- 9.0}^{+ 7.9}$ & $  15.3\pm  2.7$ & $ -0.30\pm   0.25$ & $  0.12\pm   0.22$ & 0 & $0.51$ & $21.9$ & $1.0$ & $21.1$ & $ 1.86$ & $20.3$ & $ 1.96$ \\
 88 &  3:38:28.3 & -35:27:11.5 & $  40.2_{-13.5}^{+12.5}$ & $  13.6\pm  4.4$ & $ -0.16\pm   0.70$ & $  0.31\pm   0.44$ & 0 & $0.59$ & $24.4$ & $1.0$ & \nodata & \nodata & $22.6$ & $ 2.02$ \\
 89 &  3:38:28.4 & -35:26:14.2 & $  21.1_{- 7.1}^{+ 6.0}$ & $   7.2\pm  2.2$ & $  0.09\pm   0.34$ & $  0.17\pm   1.17$ & 0 & $1.38$ & $23.3$ & $0.7$ & \nodata & \nodata & $23.1$ & $ 4.05$ \\
 90 &  3:38:28.6 & -35:27:24.7 & $  36.1_{-11.1}^{+10.0}$ & $   8.9\pm  2.6$ & $  0.03\pm   0.39$ & $ -0.27\pm   0.51$ & 0 & $0.12$ & $23.5$ & $1.0$ & \nodata & \nodata & $21.5$ & $ 2.19$ \\
 91 &  3:38:28.7 & -35:27:55.7 & $  16.1_{- 6.4}^{+ 5.3}$ & $   3.7\pm  1.3$ & $  0.24\pm   0.62$ & $  0.21\pm   0.63$ & 0 & $1.50$ & $23.8$ & $1.0$ & \nodata & \nodata & \nodata & \nodata \\
 92 &  3:38:28.8 & -35:26:18.5 & $  33.5_{- 8.1}^{+ 7.1}$ & $   9.5\pm  2.2$ & $  0.01\pm   0.28$ & $ -0.23\pm   0.43$ & 0 & $0.73$ & $23.5$ & $1.0$ & $22.6$ & $ 2.25$ & $21.7$ & $ 2.14$ \\
 93 &  3:38:28.9 & -35:29:25.6 & $  17.0_{- 6.1}^{+ 5.0}$ & $   4.5\pm  1.5$ & $  0.69\pm   0.55$ & $  0.69\pm   0.69$ & 0 & $0.36$ & $22.0$ & $1.0$ & $21.2$ & $ 1.63$ & \nodata & \nodata \\
 94 &  3:38:29.0 & -35:26: 3.1 & $ 100.0_{-11.6}^{+10.5}$ & $  26.1\pm  2.9$ & $ -0.10\pm   0.15$ & $  0.09\pm   0.15$ & 0 & $0.85$ & $24.8$ & $0.4$ & \nodata & \nodata & $23.2$ & $ 1.81$ \\
 95 &  3:38:29.0 & -35:26:30.0 & $  21.4_{- 9.8}^{+ 8.7}$ & $   7.4\pm  3.2$ & $ -0.49\pm   0.56$ & $ -0.12\pm   0.54$ & 0 & $1.83$ & $23.3$ & $1.0$ & \nodata & \nodata & \nodata & \nodata \\
 96 &  3:38:29.0 & -35:27: 1.7 & $ 700.9_{-45.3}^{+44.3}$ & $ 201.0\pm 12.8$ & $ -0.15\pm   0.07$ & $ -0.70\pm   0.12$ & 0 & $0.59$ & $15.5$ & $0.0$ & \nodata & \nodata & \nodata & \nodata \\
 97 &  3:38:29.1 & -35:26:38.0 & $  46.8_{-11.9}^{+10.9}$ & $  13.6\pm  3.3$ & $ -0.49\pm   0.38$ & $ -0.18\pm   0.29$ & 0 & $0.51$ & $22.6$ & $1.0$ & \nodata & \nodata & $20.7$ & $ 1.91$ \\
 98 &  3:38:29.2 & -35:25:53.5 & $  24.3_{- 7.0}^{+ 5.9}$ & $   5.9\pm  1.6$ & $ -0.38\pm   0.35$ & $ -0.83\pm   0.52$ & 0 & $0.89$ & $22.2$ & $1.0$ & $21.4$ & $ 1.92$ & $20.5$ & $ 2.01$ \\
 99 &  3:38:29.2 & -35:26:43.3 & $  35.1_{-11.8}^{+10.7}$ & $  10.7\pm  3.4$ & $ -0.50\pm   0.56$ & $ -0.04\pm   0.35$ & 0 & $1.46$ & $24.9$ & $0.0$ & \nodata & \nodata & \nodata & \nodata \\
100 &  3:38:29.2 & -35:28: 8.8 & $  31.1_{- 7.6}^{+ 6.5}$ & $   9.1\pm  2.1$ & $ -0.81\pm   0.28$ & $ -1.08\pm   0.46$ & 0 & $0.20$ & $18.1$ & $0.0$ & \nodata & \nodata & \nodata & \nodata \\
101 &  3:38:29.3 & -35:25:36.2 & $   8.5_{- 5.5}^{+ 4.4}$ & $   3.0\pm  1.8$ & $ 10.06\pm \infty$ & $  1.05\pm   0.74$ & 0 & $0.89$ & $24.2$ & $1.0$ & $23.2$ & $ 2.39$ & $22.3$ & $ 2.19$ \\
102 &  3:38:29.4 & -35:27: 7.1 & $ 100.6_{-24.5}^{+23.4}$ & $  35.1\pm  8.4$ & $ -0.33\pm   0.35$ & $ -0.15\pm   0.29$ & 1 & $0.37$ & $23.6$ & $1.0$ & \nodata & \nodata & $21.6$ & $ 2.43$ \\
103 &  3:38:29.4 & -35:27:30.0 & $  25.8_{- 9.2}^{+ 8.2}$ & $   6.6\pm  2.2$ & $ -0.15\pm   0.44$ & $ -0.81\pm   0.81$ & 0 & $1.90$ & $24.8$ & $1.0$ & \nodata & \nodata & \nodata & \nodata \\
104 &  3:38:29.6 & -35:27:41.2 & $  40.0_{- 8.8}^{+ 7.8}$ & $  10.0\pm  2.1$ & $ -0.28\pm   0.25$ & $ -0.53\pm   0.32$ & 1 & $0.32$ & $23.1$ & $1.0$ & \nodata & \nodata & $21.4$ & $ 1.85$ \\
105 &  3:38:29.6 & -35:31:16.0 & $  40.1_{- 8.2}^{+ 7.1}$ & $  11.3\pm  2.2$ & $ -0.42\pm   0.26$ & $ -0.40\pm   0.26$ & 0 & $0.68$ & $24.8$ & $1.0$ & \nodata & \nodata & \nodata & \nodata \\
106 &  3:38:29.7 & -35:25: 5.1 & $ 182.7_{-15.0}^{+14.0}$ & $  54.9\pm  4.4$ & $ -0.09\pm   0.09$ & $ -0.13\pm   0.11$ & 0 & $0.35$ & $23.6$ & $1.0$ & $22.7$ & $ 1.97$ & \nodata & \nodata \\
107 &  3:38:29.9 & -35:27:48.8 & $  93.1_{-11.3}^{+10.3}$ & $  23.8\pm  2.8$ & $ -0.11\pm   0.15$ & $  0.03\pm   0.15$ & 0 & $0.33$ & $22.4$ & $1.0$ & \nodata & \nodata & $20.5$ & $ 2.18$ \\
108 &  3:38:30.1 & -35:26:56.2 & $  21.7_{-12.4}^{+11.3}$ & $   6.8\pm  3.7$ & $ -0.61\pm   0.96$ & $ -0.36\pm   0.70$ & 0 & $0.45$ & $23.1$ & $1.0$ & \nodata & \nodata & $21.3$ & $ 2.16$ \\
109 &  3:38:30.2 & -35:25: 8.1 & $  34.4_{- 7.9}^{+ 6.8}$ & $   9.8\pm  2.1$ & $ -0.00\pm   0.26$ & $ -0.41\pm   0.36$ & 0 & $0.12$ & $21.9$ & $1.0$ & $21.1$ & $ 2.04$ & \nodata & \nodata \\
110 &  3:38:30.2 & -35:26:34.3 & $  25.4_{- 9.1}^{+ 8.0}$ & $   5.5\pm  1.8$ & $ -0.10\pm   0.38$ & $ -0.77\pm   0.66$ & 0 & $0.78$ & $24.1$ & $1.0$ & \nodata & \nodata & $22.4$ & $ 2.08$ \\
111 &  3:38:30.3 & -35:30:29.6 & $  44.3_{- 8.4}^{+ 7.3}$ & $  11.6\pm  2.1$ & $ -0.04\pm   0.24$ & $  0.09\pm   0.24$ & 0 & $0.28$ & $25.1$ & $1.0$ & \nodata & \nodata & \nodata & \nodata \\
112 &  3:38:30.3 & -35:31:36.9 & $  23.8_{- 6.8}^{+ 5.7}$ & $   7.8\pm  2.0$ & $ -0.00\pm   0.30$ & $ -0.86\pm   0.60$ & 1 & $0.23$ & $23.5$ & $1.0$ & $22.6$ & $ 2.08$ & \nodata & \nodata \\
113 &  3:38:30.4 & -35:24:30.8 & $  33.0_{- 7.7}^{+ 6.6}$ & $  13.0\pm  2.8$ & $ -0.29\pm   0.30$ & $  0.25\pm   0.31$ & 1 & $0.79$ & $26.6$ & $1.0$ & \nodata & \nodata & \nodata & \nodata \\
114 &  3:38:30.4 & -35:27:26.0 & $  14.9_{- 8.1}^{+ 7.1}$ & $   3.8\pm  1.9$ & $  1.03\pm   1.40$ & $  1.01\pm   0.62$ & 0 & $0.65$ & $23.3$ & $1.0$ & \nodata & \nodata & $21.6$ & $ 1.81$ \\
115 &  3:38:30.7 & -35:26:56.7 & $  23.0_{- 8.8}^{+ 7.8}$ & $   5.5\pm  2.0$ & $ -0.13\pm   0.64$ & $  0.22\pm   0.45$ & 0 & $1.42$ & $24.2$ & $1.0$ & \nodata & \nodata & \nodata & \nodata \\
116 &  3:38:30.7 & -35:27: 3.4 & $  25.6_{- 9.8}^{+ 8.7}$ & $   7.5\pm  2.7$ & $ -0.73\pm   0.51$ & $ -0.12\pm   0.36$ & 0 & $0.43$ & $23.9$ & $1.0$ & \nodata & \nodata & $22.6$ & $ 1.42$ \\
117 &  3:38:30.8 & -35:26:60.0 & $  21.5_{- 8.9}^{+ 7.8}$ & $   5.7\pm  2.2$ & $ -0.32\pm   0.53$ & $ -0.17\pm   0.44$ & 0 & $0.22$ & $23.8$ & $1.0$ & \nodata & \nodata & $22.4$ & $ 1.59$ \\
118 &  3:38:30.9 & -35:27:47.7 & $  27.3_{- 7.4}^{+ 6.4}$ & $   7.4\pm  1.9$ & $ -0.08\pm   0.43$ & $  0.39\pm   0.32$ & 0 & $0.23$ & $22.8$ & $1.0$ & \nodata & \nodata & $21.0$ & $ 2.19$ \\
119 &  3:38:31.1 & -35:25:54.8 & $  21.5_{- 7.0}^{+ 5.9}$ & $   5.9\pm  1.8$ & $ -0.47\pm   0.43$ & $ -0.09\pm   0.34$ & 0 & $3.07$ & $24.1$ & $1.0$ & \nodata & \nodata & \nodata & \nodata \\
120 &  3:38:31.2 & -35:27:39.8 & $  23.4_{- 7.0}^{+ 5.9}$ & $   6.4\pm  1.8$ & $ -0.40\pm   0.33$ & $ -0.87\pm   0.51$ & 0 & $0.30$ & $21.8$ & $0.9$ & \nodata & \nodata & $20.2$ & $ 2.15$ \\
121 &  3:38:31.2 & -35:27:48.3 & $  50.2_{- 8.9}^{+ 7.8}$ & $  17.2\pm  2.9$ & $  0.02\pm   0.24$ & $  0.49\pm   0.25$ & 0 & $0.28$ & $22.3$ & $1.0$ & $21.4$ & $ 2.03$ & $20.4$ & $ 2.23$ \\
122 &  3:38:31.2 & -35:30: 3.7 & $  84.6_{-10.6}^{+ 9.6}$ & $  22.4\pm  2.7$ & $  0.27\pm   0.16$ & $  0.38\pm   0.19$ & 0 & $0.45$ & $25.0$ & $0.0$ & $23.9$ & $ 0.69$ & \nodata & \nodata \\
123 &  3:38:31.3 & -35:24:12.1 & $ 132.8_{-13.1}^{+12.1}$ & $  37.8\pm  3.6$ & $ -0.06\pm   0.12$ & $ -0.02\pm   0.13$ & 0 & $0.07$ & $22.9$ & $1.0$ & $22.0$ & $ 1.66$ & \nodata & \nodata \\
124 &  3:38:31.3 & -35:24:53.4 & $  38.4_{- 8.1}^{+ 7.0}$ & $  17.5\pm  3.4$ & $  0.87\pm   0.68$ & $  0.99\pm   0.29$ & 0 & $0.46$ & $25.6$ & $0.4$ & \nodata & \nodata & \nodata & \nodata \\
125 &  3:38:31.5 & -35:26:50.5 & $  25.6_{- 8.1}^{+ 7.0}$ & $   8.0\pm  2.3$ & $ -0.43\pm   0.53$ & $  0.17\pm   0.32$ & 0 & $4.00$ & $25.4$ & $1.0$ & \nodata & \nodata & \nodata & \nodata \\
126 &  3:38:31.7 & -35:26: 1.1 & $  87.8_{-11.1}^{+10.1}$ & $  21.7\pm  2.6$ & $ -0.17\pm   0.15$ & $ -0.35\pm   0.18$ & 0 & $0.47$ & $22.2$ & $1.0$ & $21.2$ & $ 2.23$ & $20.3$ & $ 2.24$ \\
127 &  3:38:31.8 & -35:26: 4.6 & $ 525.3_{-24.2}^{+23.2}$ & $ 144.1\pm  6.5$ & $ -0.41\pm   0.05$ & $ -0.81\pm   0.08$ & 1 & $0.44$ & $22.4$ & $1.0$ & $21.4$ & $ 2.24$ & $20.7$ & $ 1.98$ \\
128 &  3:38:31.8 & -35:26:45.2 & $  39.2_{- 8.9}^{+ 7.8}$ & $  12.7\pm  2.7$ & $ -0.14\pm   0.30$ & $  0.17\pm   0.28$ & 0 & $0.54$ & $22.4$ & $1.0$ & \nodata & \nodata & $20.5$ & $ 2.26$ \\
129 &  3:38:31.8 & -35:27: 3.8 & $  47.8_{- 9.5}^{+ 8.5}$ & $  13.0\pm  2.5$ & $ -0.12\pm   0.27$ & $  0.12\pm   0.24$ & 1 & $1.58$ & $24.3$ & $1.0$ & \nodata & \nodata & \nodata & \nodata \\
130 &  3:38:31.8 & -35:27:45.2 & $  31.3_{- 7.5}^{+ 6.4}$ & $   7.4\pm  1.7$ & $  0.63\pm   0.41$ & $  0.72\pm   0.41$ & 0 & $0.29$ & $25.0$ & $1.0$ & \nodata & \nodata & $23.1$ & $ 1.94$ \\
131 &  3:38:31.8 & -35:30:59.4 & $ 107.2_{-11.8}^{+10.7}$ & $  30.6\pm  3.2$ & $ -0.52\pm   0.13$ & $ -0.77\pm   0.16$ & 1 & $0.30$ & $22.8$ & $1.0$ & $21.9$ & $ 2.07$ & \nodata & \nodata \\
132 &  3:38:31.9 & -35:26:49.7 & $ 132.7_{-13.4}^{+12.4}$ & $  40.8\pm  4.0$ & $ -0.41\pm   0.12$ & $ -0.58\pm   0.15$ & 0 & $0.56$ & $22.1$ & $1.0$ & \nodata & \nodata & $21.0$ & $ 1.02$ \\
133 &  3:38:32.1 & -35:28:13.1 & $  26.7_{- 7.2}^{+ 6.1}$ & $   6.5\pm  1.6$ & $ -0.03\pm   0.34$ & $  0.04\pm   0.35$ & 1 & $0.19$ & $21.0$ & $1.0$ & $20.2$ & $ 1.66$ & $19.5$ & $ 1.84$ \\
134 &  3:38:32.2 & -35:27: 6.3 & $  43.5_{- 8.9}^{+ 7.9}$ & $  11.5\pm  2.2$ & $ -0.12\pm   0.27$ & $  0.13\pm   0.24$ & 0 & $0.57$ & $23.5$ & $1.0$ & \nodata & \nodata & $21.6$ & $ 2.16$ \\
135 &  3:38:32.3 & -35:26:46.5 & $  69.0_{-10.4}^{+ 9.3}$ & $  24.2\pm  3.5$ & $ -0.61\pm   0.18$ & $ -0.80\pm   0.23$ & 0 & $1.32$ & $26.1$ & $1.0$ & \nodata & \nodata & \nodata & \nodata \\
136 &  3:38:32.3 & -35:27:11.0 & $ 110.2_{-12.3}^{+11.3}$ & $  28.7\pm  3.1$ & $ -0.15\pm   0.14$ & $ -0.12\pm   0.15$ & 0 & $1.14$ & $22.7$ & $1.0$ & \nodata & \nodata & $25.3$ & $ 1.42$ \\
137 &  3:38:32.4 & -35:27: 2.6 & $  98.5_{-11.9}^{+10.9}$ & $  26.8\pm  3.1$ & $  0.05\pm   0.14$ & $  0.11\pm   0.17$ & 0 & $0.46$ & $24.2$ & $1.0$ & \nodata & \nodata & $22.7$ & $ 1.72$ \\
138 &  3:38:32.4 & -35:27:29.6 & $  64.9_{- 9.9}^{+ 8.9}$ & $  17.1\pm  2.5$ & $  0.01\pm   0.18$ & $ -0.00\pm   0.21$ & 0 & $0.36$ & $23.0$ & $1.0$ & \nodata & \nodata & $23.2$ & $ 4.93$ \\
139 &  3:38:32.4 & -35:27:35.0 & $  26.8_{- 7.2}^{+ 6.1}$ & $   7.3\pm  1.8$ & $  0.04\pm   0.39$ & $  0.41\pm   0.34$ & 0 & $0.42$ & $26.3$ & $0.0$ & \nodata & \nodata & $21.3$ & $ 1.95$ \\
140 &  3:38:32.5 & -35:24:41.1 & $  32.8_{- 7.6}^{+ 6.6}$ & $  11.2\pm  2.4$ & $ -0.16\pm   0.31$ & $  0.39\pm   0.31$ & 0 & $0.47$ & $22.6$ & $1.0$ & $21.8$ & $ 1.91$ & \nodata & \nodata \\
141 &  3:38:32.6 & -35:27: 5.9 & $ 856.3_{-30.5}^{+29.5}$ & $ 238.3\pm  8.4$ & $ -0.08\pm   0.04$ & $  0.10\pm   0.05$ & 0 & $0.43$ & $21.7$ & $1.0$ & $21.0$ & $ 2.12$ & $19.9$ & $ 2.24$ \\
142 &  3:38:32.6 & -35:27:43.0 & $  18.6_{- 6.5}^{+ 5.4}$ & $   4.2\pm  1.4$ & $  0.00\pm   0.38$ & $ -0.97\pm   0.87$ & 0 & $2.95$ & $25.3$ & $0.9$ & \nodata & \nodata & \nodata & \nodata \\
143 &  3:38:32.7 & -35:26:53.4 & $  39.2_{- 8.5}^{+ 7.5}$ & $  10.0\pm  2.0$ & $ -0.18\pm   0.28$ & $ -0.00\pm   0.26$ & 0 & $0.34$ & $23.6$ & $1.0$ & \nodata & \nodata & $22.2$ & $ 1.51$ \\
144 &  3:38:32.8 & -35:26:22.4 & $  16.2_{- 6.6}^{+ 5.5}$ & $   3.3\pm  1.2$ & $  1.07\pm   0.61$ & $  1.28\pm   1.92$ & 0 & $0.89$ & $26.1$ & $1.0$ & \nodata & \nodata & $24.2$ & $ 2.39$ \\
145 &  3:38:32.8 & -35:26:59.0 & $ 157.0_{-14.2}^{+13.1}$ & $  39.6\pm  3.4$ & $  0.13\pm   0.11$ & $  0.17\pm   0.13$ & 0 & $0.47$ & $22.9$ & $1.0$ & $22.2$ & $ 1.29$ & $21.5$ & $ 1.55$ \\
146 &  3:38:32.9 & -35:32: 5.0 & $  34.0_{- 7.7}^{+ 6.6}$ & $  17.3\pm  3.6$ & $ -0.01\pm   0.29$ & $  0.46\pm   0.37$ & 0 & $0.90$ & $22.0$ & $0.9$ & \nodata & \nodata & \nodata & \nodata \\
147 &  3:38:33.0 & -35:28:29.0 & $  46.6_{- 8.6}^{+ 7.5}$ & $  12.9\pm  2.2$ & $ -0.21\pm   0.22$ & $ -0.09\pm   0.23$ & 0 & $0.27$ & $22.4$ & $1.0$ & $21.6$ & $ 2.13$ & $20.5$ & $ 2.23$ \\
148 &  3:38:33.1 & -35:26: 2.1 & $  45.0_{- 8.7}^{+ 7.6}$ & $  11.0\pm  2.0$ & $  0.03\pm   0.25$ & $  0.05\pm   0.26$ & 0 & $1.51$ & $26.1$ & $0.4$ & \nodata & \nodata & $25.2$ & $ 0.64$ \\
149 &  3:38:33.1 & -35:26:52.5 & $  26.8_{- 7.5}^{+ 6.4}$ & $   5.3\pm  1.4$ & $  0.80\pm   0.43$ & $  0.65\pm   0.59$ & 1 & $0.43$ & $21.8$ & $1.0$ & \nodata & \nodata & $20.2$ & $ 1.66$ \\
150 &  3:38:33.1 & -35:27:32.1 & $ 229.5_{-16.5}^{+15.5}$ & $  65.0\pm  4.5$ & $ -0.32\pm   0.09$ & $ -0.21\pm   0.09$ & 0 & $0.33$ & $23.4$ & $1.0$ & \nodata & \nodata & $20.4$ & $ 1.21$ \\
151 &  3:38:33.1 & -35:28:14.9 & $  17.6_{- 6.3}^{+ 5.2}$ & $   5.4\pm  1.8$ & $  0.09\pm   0.35$ & $ -0.17\pm  \infty$ & 0 & $2.78$ & $27.7$ & $0.6$ & \nodata & \nodata & \nodata & \nodata \\
152 &  3:38:33.1 & -35:31: 3.5 & $  43.9_{- 8.2}^{+ 7.1}$ & $  11.8\pm  2.1$ & $ -0.22\pm   0.22$ & $ -0.45\pm   0.27$ & 0 & $0.12$ & $22.6$ & $1.0$ & $21.9$ & $ 1.49$ & \nodata & \nodata \\
153 &  3:38:33.2 & -35:25:54.2 & $  48.7_{- 8.8}^{+ 7.7}$ & $  14.4\pm  2.4$ & $ -0.29\pm   0.23$ & $ -0.01\pm   0.22$ & 1 & $0.61$ & $23.2$ & $1.0$ & $22.5$ & $ 2.02$ & $21.4$ & $ 2.14$ \\
154 &  3:38:33.4 & -35:23: 2.8 & $ 266.8_{-17.9}^{+16.9}$ & $  71.2\pm  4.6$ & $  0.53\pm   0.10$ & $  0.65\pm   0.12$ & 0 & $0.47$ & $27.5$ & $0.1$ & \nodata & \nodata & \nodata & \nodata \\
155 &  3:38:33.4 & -35:32:29.5 & $  32.8_{- 7.5}^{+ 6.4}$ & $  13.8\pm  2.9$ & $ -0.45\pm   0.31$ & $ -0.60\pm   0.35$ & 0 & $0.60$ & $21.7$ & $1.0$ & $21.0$ & $ 1.94$ & \nodata & \nodata \\
156 &  3:38:33.5 & -35:26:46.4 & $  11.0_{- 6.2}^{+ 5.1}$ & $   1.6\pm  0.8$ & $  1.33\pm  \infty$ & $ -8.34\pm\infty$ & 0 & $0.51$ & $24.5$ & $1.0$ & \nodata & \nodata & $22.6$ & $ 2.01$ \\
157 &  3:38:33.6 & -35:23:24.2 & $  54.4_{- 9.7}^{+ 8.7}$ & $  17.0\pm  2.9$ & $  0.11\pm   0.25$ & $  0.51\pm   0.25$ & 0 & $0.31$ & $22.7$ & $1.0$ & $22.0$ & $ 1.33$ & \nodata & \nodata \\
158 &  3:38:33.8 & -35:25: 9.8 & $  20.2_{- 6.4}^{+ 5.3}$ & $   6.2\pm  1.8$ & $ -0.31\pm   0.41$ & $ -0.07\pm   0.36$ & 1 & $2.03$ & $27.5$ & $0.0$ & \nodata & \nodata & \nodata & \nodata \\
159 &  3:38:33.8 & -35:25:57.4 & $  84.2_{-10.8}^{+ 9.8}$ & $  21.6\pm  2.6$ & $ -0.28\pm   0.15$ & $ -0.55\pm   0.19$ & 0 & $0.41$ & $20.5$ & $1.0$ & $19.8$ & $ 1.57$ & $19.1$ & $ 1.79$ \\
160 &  3:38:33.8 & -35:26:58.9 & $  67.0_{-10.1}^{+ 9.0}$ & $  16.5\pm  2.4$ & $ -0.00\pm   0.18$ & $ -0.11\pm   0.21$ & 0 & $0.37$ & $22.4$ & $1.0$ & $21.6$ & $ 2.09$ & $20.6$ & $ 2.16$ \\
161 &  3:38:34.1 & -35:27:57.6 & $  11.1_{- 5.5}^{+ 4.4}$ & $   3.2\pm  1.4$ & $ -0.13\pm   0.60$ & $ -0.03\pm   0.60$ & 0 & $3.89$ & $25.2$ & $0.9$ & \nodata & \nodata & \nodata & \nodata \\
162 &  3:38:34.2 & -35:29:51.7 & $  44.2_{- 8.1}^{+ 7.1}$ & $  15.4\pm  2.6$ & $  0.05\pm   0.25$ & $  0.47\pm   0.26$ & 0 & $0.37$ & $25.3$ & $0.6$ & $24.1$ & $ 0.43$ & \nodata & \nodata \\
163 &  3:38:34.3 & -35:30:14.1 & $  22.7_{- 6.6}^{+ 5.5}$ & $   7.3\pm  1.9$ & $ -0.49\pm   0.42$ & $  0.08\pm   0.31$ & 0 & $0.39$ & $24.4$ & $1.0$ & \nodata & \nodata & \nodata & \nodata \\
164 &  3:38:34.7 & -35:28:43.7 & $  19.8_{- 6.4}^{+ 5.3}$ & $   5.2\pm  1.5$ & $ -0.26\pm   0.33$ & $ -1.05\pm   0.66$ & 1 & $0.90$ & $24.3$ & $0.2$ & \nodata & \nodata & $23.0$ & $ 1.73$ \\
165 &  3:38:35.0 & -35:26:55.6 & $  15.5_{- 6.2}^{+ 5.1}$ & $   6.8\pm  2.5$ & $ -0.61\pm   0.65$ & $  0.14\pm   0.41$ & 0 & $0.25$ & $22.8$ & $1.0$ & $22.0$ & $ 2.03$ & $21.0$ & $ 2.13$ \\
166 &  3:38:35.1 & -35:25:29.3 & $   6.9_{- 4.9}^{+ 3.7}$ & $   2.6\pm  1.6$ & $ -0.84\pm   1.09$ & $ -0.15\pm   0.65$ & 0 & $2.22$ & $27.3$ & $0.5$ & \nodata & \nodata & \nodata & \nodata \\
167 &  3:38:35.2 & -35:28:24.7 & $  37.6_{- 7.8}^{+ 6.7}$ & $   9.7\pm  1.9$ & $ -0.19\pm   0.24$ & $ -0.46\pm   0.31$ & 0 & $0.26$ & $22.3$ & $1.0$ & $21.4$ & $ 1.81$ & $20.7$ & $ 1.90$ \\
168 &  3:38:35.4 & -35:29: 5.8 & $  71.6_{- 9.9}^{+ 8.9}$ & $  20.9\pm  2.7$ & $  0.54\pm   0.20$ & $  0.66\pm   0.26$ & 0 & $0.26$ & $22.6$ & $1.0$ & $21.5$ & $ 0.36$ & \nodata & \nodata \\
169 &  3:38:35.7 & -35:30:24.1 & $  32.9_{- 7.4}^{+ 6.3}$ & $  10.8\pm  2.2$ & $  0.04\pm   0.28$ & $  0.39\pm   0.33$ & 0 & $0.57$ & $22.8$ & $0.9$ & $22.0$ & $ 1.54$ & \nodata & \nodata \\
170 &  3:38:35.7 & -35:31: 2.4 & $  32.6_{- 7.2}^{+ 6.1}$ & $  13.3\pm  2.7$ & $ -0.13\pm   0.28$ & $ -0.19\pm   0.30$ & 0 & $0.24$ & $21.2$ & $1.0$ & $20.5$ & $ 1.57$ & \nodata & \nodata \\
171 &  3:38:35.9 & -35:26:16.7 & $  58.0_{- 9.3}^{+ 8.3}$ & $  16.0\pm  2.4$ & $  0.09\pm   0.20$ & $  0.12\pm   0.24$ & 0 & $0.50$ & $21.8$ & $1.0$ & \nodata & \nodata & $20.2$ & $ 2.07$ \\
172 &  3:38:35.9 & -35:28:32.0 & $  25.6_{- 6.8}^{+ 5.7}$ & $   6.9\pm  1.7$ & $ -0.39\pm   0.33$ & $ -0.29\pm   0.32$ & 0 & $0.20$ & $27.5$ & $0.9$ & \nodata & \nodata & \nodata & \nodata \\
173 &  3:38:36.2 & -35:26:26.1 & $  54.3_{- 9.1}^{+ 8.0}$ & $  15.7\pm  2.5$ & $ -0.58\pm   0.21$ & $ -0.63\pm   0.22$ & 0 & $0.26$ & $22.1$ & $1.0$ & $22.0$ & $ 0.56$ & $20.7$ & $ 1.26$ \\
174 &  3:38:36.3 & -35:27: 9.2 & $  71.6_{-10.0}^{+ 8.9}$ & $  19.1\pm  2.5$ & $ -0.20\pm   0.16$ & $ -0.49\pm   0.21$ & 0 & $0.31$ & $22.3$ & $1.0$ & $21.5$ & $ 1.96$ & $20.5$ & $ 2.03$ \\
175 &  3:38:36.3 & -35:28: 9.9 & $  79.6_{-10.4}^{+ 9.3}$ & $  19.9\pm  2.5$ & $  0.36\pm   0.18$ & $  0.54\pm   0.20$ & 0 & $0.18$ & $22.7$ & $0.0$ & \nodata & \nodata & $21.8$ & $ 3.04$ \\
176 &  3:38:36.8 & -35:27:47.5 & $ 151.5_{-13.7}^{+12.6}$ & $  46.1\pm  4.0$ & $ -0.58\pm   0.12$ & $ -0.50\pm   0.12$ & 1 & $0.19$ & $23.0$ & $1.0$ & $22.9$ & $ 0.69$ & $21.8$ & $ 1.03$ \\
177 &  3:38:36.9 & -35:25:42.7 & $  21.5_{- 6.3}^{+ 5.2}$ & $   4.6\pm  1.2$ & $  0.08\pm   0.39$ & $ -0.05\pm   0.43$ & 0 & $0.45$ & $21.6$ & $0.9$ & \nodata & \nodata & $19.9$ & $ 2.28$ \\
178 &  3:38:37.1 & -35:25: 6.0 & $  17.7_{- 5.9}^{+ 4.8}$ & $   4.5\pm  1.4$ & $ -0.58\pm   0.38$ & $ -0.70\pm   0.45$ & 0 & $0.34$ & $22.3$ & $1.0$ & \nodata & \nodata & \nodata & \nodata \\
179 &  3:38:37.2 & -35:28:13.3 & $  20.1_{- 6.2}^{+ 5.1}$ & $   5.7\pm  1.6$ & $ -0.69\pm   0.41$ & $ -0.44\pm   0.34$ & 0 & $0.37$ & $21.6$ & $1.0$ & $20.8$ & $ 1.89$ & $19.9$ & $ 2.08$ \\
180 &  3:38:37.3 & -35:27:10.2 & $  18.9_{- 6.2}^{+ 5.1}$ & $   5.1\pm  1.5$ & $ -0.47\pm   0.43$ & $ -0.40\pm   0.41$ & 0 & $1.75$ & $22.8$ & $0.0$ & \nodata & \nodata & \nodata & \nodata \\
181 &  3:38:38.0 & -35:26:59.5 & $  21.7_{- 6.5}^{+ 5.4}$ & $   6.7\pm  1.8$ & $ -0.61\pm   0.44$ & $ -0.25\pm   0.33$ & 0 & $0.25$ & $21.6$ & $1.0$ & $20.9$ & $ 1.70$ & $20.1$ & $ 1.88$ \\
182 &  3:38:38.2 & -35:28: 4.7 & $  23.0_{- 6.5}^{+ 5.4}$ & $   9.1\pm  2.4$ & $ -0.32\pm   0.33$ & $  0.18\pm   0.40$ & 0 & $4.35$ & $22.2$ & $1.0$ & \nodata & \nodata & \nodata & \nodata \\
183 &  3:38:38.4 & -35:29:16.1 & $  36.4_{- 7.6}^{+ 6.5}$ & $  10.0\pm  2.0$ & $ -0.20\pm   0.26$ & $  0.03\pm   0.26$ & 0 & $4.78$ & $27.9$ & $0.7$ & \nodata & \nodata & \nodata & \nodata \\
184 &  3:38:38.5 & -35:29:26.3 & $  31.3_{- 7.2}^{+ 6.2}$ & $   9.9\pm  2.1$ & $ -0.27\pm   0.29$ & $ -0.00\pm   0.29$ & 0 & $0.24$ & $21.8$ & $1.0$ & $21.0$ & $ 1.83$ & \nodata & \nodata \\
185 &  3:38:38.6 & -35:27:28.5 & $  68.2_{- 9.7}^{+ 8.7}$ & $  17.1\pm  2.3$ & $ -0.04\pm   0.17$ & $ -0.12\pm   0.20$ & 0 & $0.16$ & $21.0$ & $1.0$ & $20.2$ & $ 1.75$ & $19.4$ & $ 1.90$ \\
186 &  3:38:38.7 & -35:26:40.6 & $  20.3_{- 6.2}^{+ 5.1}$ & $   6.7\pm  1.9$ & $ -0.18\pm   0.37$ & $ -0.21\pm   0.46$ & 1 & $0.20$ & $24.4$ & $1.0$ & $21.9$ & $ 1.95$ & $20.9$ & $ 2.15$ \\
187 &  3:38:38.7 & -35:28: 5.0 & $  13.0_{- 5.5}^{+ 4.3}$ & $   3.6\pm  1.4$ & $ -0.36\pm   0.49$ & $ -0.38\pm   0.53$ & 0 & $0.42$ & $21.4$ & $0.9$ & \nodata & \nodata & \nodata & \nodata \\
188 &  3:38:38.8 & -35:25:43.1 & $  18.5_{- 5.9}^{+ 4.8}$ & $   5.8\pm  1.7$ & $ -0.15\pm   0.48$ & $  0.44\pm   0.39$ & 0 & $0.52$ & $21.8$ & $0.9$ & $21.0$ & $ 1.87$ & $20.2$ & $ 2.06$ \\
189 &  3:38:38.8 & -35:25:55.2 & $ 133.3_{-12.8}^{+11.7}$ & $  34.1\pm  3.1$ & $ -0.13\pm   0.12$ & $ -0.05\pm   0.12$ & 0 & $0.31$ & $21.4$ & $0.9$ & $20.5$ & $ 1.91$ & $19.7$ & $ 2.09$ \\
190 &  3:38:38.8 & -35:27:21.6 & $  22.0_{- 6.4}^{+ 5.3}$ & $   6.7\pm  1.8$ & $ -0.23\pm   0.40$ & $ -0.00\pm   0.37$ & 0 & $0.15$ & $22.1$ & $0.9$ & $21.3$ & $ 1.50$ & $20.9$ & $ 1.67$ \\
191 &  3:38:39.3 & -35:26:35.0 & $  11.9_{- 5.2}^{+ 4.1}$ & $   3.2\pm  1.2$ & $ -0.44\pm   0.48$ & $ -0.74\pm   0.64$ & 0 & $0.46$ & $22.4$ & $1.0$ & $21.6$ & $ 1.72$ & $20.9$ & $ 1.80$ \\
192 &  3:38:39.3 & -35:30:19.8 & $   6.9_{- 4.9}^{+ 3.7}$ & $   1.2\pm  0.8$ & $ -0.74\pm   1.01$ & $ -0.76\pm   0.93$ & 0 & $0.14$ & $26.2$ & $0.0$ & \nodata & \nodata & \nodata & \nodata \\
193 &  3:38:39.8 & -35:29:54.8 & $  16.1_{- 5.8}^{+ 4.7}$ & $   3.5\pm  1.1$ & $  0.03\pm   0.35$ & $ -1.44\pm   2.03$ & 0 & $0.21$ & $23.2$ & $1.0$ & $22.4$ & $ 1.65$ & \nodata & \nodata \\
194 &  3:38:39.9 & -35:27:33.1 & $  22.3_{- 6.4}^{+ 5.3}$ & $   5.1\pm  1.3$ & $  0.06\pm   0.36$ & $ -0.03\pm   0.39$ & 0 & $2.68$ & $25.6$ & $0.0$ & \nodata & \nodata & \nodata & \nodata \\
195 &  3:38:40.4 & -35:28:47.6 & $  20.4_{- 6.4}^{+ 5.3}$ & $   5.2\pm  1.5$ & $  0.07\pm   0.42$ & $  0.30\pm   0.40$ & 0 & $0.18$ & $21.5$ & $1.0$ & $20.7$ & $ 1.89$ & \nodata & \nodata \\
196 &  3:38:40.5 & -35:26:47.5 & $  15.3_{- 5.6}^{+ 4.4}$ & $   3.1\pm  1.0$ & $  0.23\pm   0.36$ & $ -0.79\pm   0.80$ & 0 & $0.28$ & $22.6$ & $1.0$ & $21.7$ & $ 1.71$ & \nodata & \nodata \\
197 &  3:38:40.8 & -35:26: 8.5 & $  29.4_{- 6.9}^{+ 5.8}$ & $   7.5\pm  1.6$ & $ -0.18\pm   0.28$ & $ -0.14\pm   0.30$ & 0 & $0.39$ & $24.6$ & $0.8$ & $21.7$ & $ 1.19$ & \nodata & \nodata \\
198 &  3:38:41.4 & -35:31:34.6 & $3919.6_{-63.7}^{+62.7}$ & $1566.3\pm 25.2$ & $ -0.37\pm   0.02$ & $ -0.40\pm   0.02$ & 0 & $0.15$ & $19.9$ & $0.9$ & $19.1$ & $-0.10$ & \nodata & \nodata \\
199 &  3:38:41.8 & -35:23: 3.7 & $  12.9_{- 5.3}^{+ 4.2}$ & $   7.9\pm  2.9$ & $ -0.54\pm   0.53$ & $ -0.79\pm   0.69$ & 0 & $1.36$ & $22.0$ & $1.0$ & \nodata & \nodata & \nodata & \nodata \\
200 &  3:38:41.9 & -35:24:43.2 & $  30.4_{- 7.2}^{+ 6.1}$ & $  10.5\pm  2.3$ & $ -0.17\pm   0.34$ & $  0.34\pm   0.30$ & 0 & $0.86$ & $22.0$ & $1.0$ & $21.1$ & $ 2.18$ & \nodata & \nodata \\
201 &  3:38:41.9 & -35:26: 1.0 & $  12.1_{- 5.2}^{+ 4.1}$ & $   2.7\pm  1.0$ & $  0.10\pm   0.51$ & $  0.25\pm   0.56$ & 0 & $0.47$ & $21.4$ & $0.9$ & $20.6$ & $ 2.08$ & \nodata & \nodata \\
202 &  3:38:42.1 & -35:26:18.9 & $  47.4_{- 8.3}^{+ 7.2}$ & $  12.3\pm  2.0$ & $ -0.07\pm   0.23$ & $  0.21\pm   0.22$ & 0 & $0.41$ & $22.0$ & $1.0$ & $21.2$ & $ 1.76$ & \nodata & \nodata \\
203 &  3:38:42.4 & -35:24: 1.0 & $ 135.0_{-13.1}^{+12.0}$ & $  34.5\pm  3.2$ & $ -0.04\pm   0.12$ & $ -0.04\pm   0.13$ & 1 & $0.75$ & $23.6$ & $1.0$ & \nodata & \nodata & \nodata & \nodata \\
204 &  3:38:42.5 & -35:27:46.8 & $   9.3_{- 4.9}^{+ 3.7}$ & $   2.4\pm  1.1$ & $ -0.29\pm   0.67$ & $ -0.40\pm   0.72$ & 0 & $1.37$ & $24.7$ & $0.0$ & \nodata & \nodata & \nodata & \nodata \\
205 &  3:38:42.6 & -35:29:12.4 & $  15.7_{- 5.7}^{+ 4.6}$ & $   5.1\pm  1.7$ & $  0.12\pm   0.44$ & $  0.43\pm   0.54$ & 0 & $0.36$ & $22.9$ & $1.0$ & $22.0$ & $ 1.77$ & \nodata & \nodata \\
206 &  3:38:43.1 & -35:23:41.7 & $  33.9_{- 7.5}^{+ 6.4}$ & $  16.2\pm  3.3$ & $  0.05\pm   0.30$ & $  0.23\pm   0.36$ & 1 & $0.41$ & $27.3$ & $0.7$ & \nodata & \nodata & \nodata & \nodata \\
207 &  3:38:43.1 & -35:28: 3.2 & $  34.5_{- 7.4}^{+ 6.3}$ & $   7.6\pm  1.5$ & $  0.29\pm   0.25$ & $ -0.06\pm   0.36$ & 0 & $0.12$ & $21.8$ & $1.0$ & $21.0$ & $ 1.53$ & \nodata & \nodata \\
208 &  3:38:43.2 & -35:27:35.9 & $  38.9_{- 7.7}^{+ 6.6}$ & $  13.4\pm  2.5$ & $ -0.44\pm   0.25$ & $  0.02\pm   0.24$ & 1 & $0.23$ & $22.3$ & $1.0$ & $21.5$ & $ 1.89$ & \nodata & \nodata \\
209 &  3:38:43.2 & -35:29:40.3 & $  29.9_{- 7.1}^{+ 6.0}$ & $   8.7\pm  1.9$ & $ -0.14\pm   0.28$ & $  0.12\pm   0.29$ & 0 & $0.59$ & $23.2$ & $1.0$ & $22.3$ & $ 1.74$ & \nodata & \nodata \\
210 &  3:38:43.2 & -35:31:25.2 & $  53.3_{- 8.9}^{+ 7.9}$ & $  22.7\pm  3.6$ & $ -0.12\pm   0.28$ & $ -0.00\pm   0.28$ & 0 & $4.12$ & $25.1$ & $0.0$ & \nodata & \nodata & \nodata & \nodata \\
211 &  3:38:43.3 & -35:24:14.7 & $ 124.0_{-12.6}^{+11.5}$ & $  35.4\pm  3.4$ & $ -0.06\pm   0.13$ & $  0.09\pm   0.14$ & 0 & $0.51$ & $23.5$ & $1.0$ & $22.6$ & $ 1.93$ & \nodata & \nodata \\
212 &  3:38:43.5 & -35:26: 4.1 & $  32.2_{- 7.2}^{+ 6.1}$ & $   9.6\pm  2.0$ & $ -0.36\pm   0.32$ & $  0.06\pm   0.26$ & 0 & $0.40$ & $23.0$ & $1.0$ & $22.1$ & $ 1.71$ & \nodata & \nodata \\
213 &  3:38:44.0 & -35:25: 4.2 & $  33.7_{- 7.5}^{+ 6.4}$ & $  10.2\pm  2.1$ & $ -0.06\pm   0.32$ & $  0.12\pm   0.30$ & 0 & $0.28$ & $22.1$ & $1.0$ & $21.3$ & $ 1.72$ & \nodata & \nodata \\
214 &  3:38:44.7 & -35:26:46.1 & $  16.9_{- 5.8}^{+ 4.7}$ & $   4.3\pm  1.3$ & $ -0.29\pm   0.38$ & $ -0.52\pm   0.47$ & 0 & $0.53$ & $22.7$ & $1.0$ & $21.8$ & $ 1.79$ & \nodata & \nodata \\
215 &  3:38:44.8 & -35:25:47.1 & $  13.3_{- 5.3}^{+ 4.2}$ & $   6.1\pm  2.2$ & $ -0.26\pm   0.54$ & $  0.31\pm   0.49$ & 0 & $0.86$ & $21.6$ & $1.0$ & $20.8$ & $ 1.87$ & \nodata & \nodata \\
216 &  3:38:45.0 & -35:27:55.8 & $  40.5_{- 7.8}^{+ 6.8}$ & $   9.6\pm  1.7$ & $  0.03\pm   0.21$ & $ -0.74\pm   0.39$ & 0 & $5.36$ & $26.0$ & $0.0$ & \nodata & \nodata & \nodata & \nodata \\
217 &  3:38:45.0 & -35:28:22.2 & $  70.4_{- 9.8}^{+ 8.7}$ & $  19.0\pm  2.5$ & $ -0.11\pm   0.17$ & $ -0.01\pm   0.18$ & 0 & $0.23$ & $21.7$ & $0.9$ & $20.8$ & $ 1.91$ & \nodata & \nodata \\
218 &  3:38:45.3 & -35:27:37.9 & $ 224.4_{-16.2}^{+15.2}$ & $  59.2\pm  4.1$ & $  0.11\pm   0.09$ & $  0.18\pm   0.10$ & 0 & $0.26$ & $25.6$ & $0.8$ & \nodata & \nodata & \nodata & \nodata \\
219 &  3:38:45.6 & -35:29:37.1 & $  21.2_{- 6.3}^{+ 5.2}$ & $   6.5\pm  1.8$ & $  0.30\pm   0.35$ & $ -0.23\pm   0.64$ & 0 & $2.97$ & $26.6$ & $0.0$ & \nodata & \nodata & \nodata & \nodata \\
220 &  3:38:46.9 & -35:29:50.4 & $  97.6_{-11.2}^{+10.2}$ & $  50.4\pm  5.5$ & $ -0.36\pm   0.17$ & $ -0.45\pm   0.19$ & 0 & $0.37$ & $24.3$ & $1.0$ & $23.2$ & $ 1.70$ & \nodata & \nodata \\
221 &  3:38:48.2 & -35:27:58.4 & $  17.1_{- 5.9}^{+ 4.8}$ & $   5.9\pm  1.9$ & $ -0.36\pm   0.39$ & $ -0.08\pm   0.47$ & 0 & $0.29$ & $21.4$ & $0.9$ & $20.6$ & $ 1.73$ & \nodata & \nodata \\
222 &  3:38:48.4 & -35:27: 4.1 & $  23.9_{- 6.7}^{+ 5.6}$ & $   7.6\pm  1.9$ & $ -0.38\pm   0.39$ & $ -0.11\pm   0.33$ & 0 & $1.27$ & $22.8$ & $1.0$ & \nodata & \nodata & \nodata & \nodata \\
223 &  3:38:48.6 & -35:29:21.3 & $  31.6_{- 7.2}^{+ 6.2}$ & $  10.8\pm  2.3$ & $  0.10\pm   0.26$ & $ -0.77\pm   0.52$ & 0 & $2.53$ & $25.6$ & $0.3$ & \nodata & \nodata & \nodata & \nodata \\
224 &  3:38:48.7 & -35:28:35.0 & $ 226.5_{-16.3}^{+15.3}$ & $  62.6\pm  4.4$ & $ -0.32\pm   0.08$ & $ -0.56\pm   0.11$ & 1 & $0.26$ & $21.5$ & $0.9$ & $20.7$ & $ 0.56$ & \nodata & \nodata \\
225 &  3:38:49.0 & -35:29:50.1 & $  60.5_{- 9.4}^{+ 8.3}$ & $  16.8\pm  2.4$ & $  0.09\pm   0.20$ & $  0.16\pm   0.21$ & 0 & $0.39$ & $26.5$ & $0.0$ & \nodata & \nodata & \nodata & \nodata \\
226 &  3:38:50.0 & -35:29: 9.3 & $  26.5_{- 7.0}^{+ 5.9}$ & $  16.4\pm  4.0$ & $ -0.46\pm   0.34$ & $  0.27\pm   0.33$ & 0 & $4.20$ & $27.7$ & $0.8$ & \nodata & \nodata & \nodata & \nodata \\
227 &  3:38:51.1 & -35:30: 8.3 & $  58.4_{- 9.2}^{+ 8.1}$ & $  16.3\pm  2.4$ & $  0.86\pm   0.28$ & $  0.90\pm   0.29$ & 1 & $4.16$ & $27.6$ & $0.8$ & \nodata & \nodata & \nodata & \nodata \\
228 &  3:38:51.6 & -35:26:44.8 & $1479.5_{-39.6}^{+38.6}$ & $ 421.3\pm 11.1$ & $ -0.10\pm   0.03$ & $ -0.11\pm   0.04$ & 0 & $0.71$ & $20.2$ & $1.0$ & \nodata & \nodata & \nodata & \nodata \\
229 &  3:38:53.8 & -35:29:47.1 & $  18.8_{- 6.4}^{+ 5.3}$ & $   6.6\pm  2.0$ & $  0.27\pm   0.46$ & $  0.60\pm   0.50$ & 0 & $0.72$ & $23.3$ & $1.0$ & $22.4$ & $ 1.79$ & \nodata & \nodata \\
230 &  3:38:56.2 & -35:27:55.1 & $  28.4_{- 7.4}^{+ 6.3}$ & $  11.6\pm  2.8$ & $ -0.22\pm   0.35$ & $ -0.29\pm   0.37$ & 0 & $8.47$ & $20.9$ & $0.0$ & $21.3$ & $ 1.46$ & \nodata & \nodata \\
\enddata
  \tablecomments{Table \ref{tbl-1} is published in its entirety in the 
  electronic edition of the {\it Astrophysical Journal}.  A portion is 
  shown here for guidance regarding its form and content.}
  \tablenotetext{}{\textbf{ID}: Source index; \textbf{RA, DEC}: USNO registered coordinates; \textbf{Net counts}: Total net counts in the 0.5-8 keV band; \textbf{Flux}: Photon flux in the 0.5-8 keV band; \textbf{HR1, HR2}: Hardness ratios computed from photon fluxex in the $0.5-1~vs~1-2$ keV and $0.5-1~vs~2-8$ keV band; \textbf{Var.}: variability flag based on either K-S probability $P>95$\% or flux difference $>3 \sigma$ between obs. \#319 and \#1472 (1=variable 0=non-variable); \textbf{X-opt sep.}: separation between the X-ray source and the closest HST $V$ band counterpart; \textbf{\textit{V}}: F606V magnitude ; \textbf{Stellar index}: SExtractor stellarity index; \textbf{\textit{T1}}: T1 ground-based magnitude; \textbf{\textit{C-T1}}: C-T1 ground-based color; \textbf{\textit{z}}: ACS $z$ magnitude; \textbf{\textit{g-z}}: ACS $g-z$ color}
\end{deluxetable}
\clearpage
\end{landscape}

\end{document}